\documentclass[
aps,
pra,
floatfix,
]{revtex4-2} 
\usepackage{amsmath,amsthm,amssymb,epsfig,graphicx,mathrsfs,amsfonts,dsfont,bbm}
\usepackage{pict2e}
\usepackage[percent]{overpic}
\usepackage{color}
\usepackage{listings}
\usepackage{caption}
\usepackage[toc,title,titletoc,header]{appendix}
\usepackage{color}
\usepackage{dcolumn}
\usepackage{bm}
\usepackage{hyperref}
\hypersetup{
    citecolor=magenta,
    colorlinks=true,
    linkcolor=blue,
    filecolor=green,      
    urlcolor=cyan,
}
\usepackage[capitalise]{cleveref}
\usepackage{subcaption}
\usepackage{enumitem}
\usepackage{mathtools}
\usepackage{tikz}
\usepackage{tikzit}

\tikzstyle{bold vertex}=[fill={rgb,255: red,252; green,156; blue,255}, draw=black, shape=circle]

\tikzstyle{bold edge}=[->, fill=none, line width=1.5pt, draw=black]
\tikzstyle{blue edge}=[fill=none, draw={rgb,255: red,125; green,184; blue,255}, ->, line width=1.5pt]
\tikzstyle{red edge}=[draw={rgb,255: red,255; green,80; blue,83}, ->, line width=1.5pt]
\tikzstyle{bold red edge}=[-, draw={rgb,255: red,255; green,80; blue,83}, line width=3pt]
\tikzstyle{bold blue}=[-, draw={rgb,255: red,125; green,184; blue,255}, line width=3pt]
\tikzstyle{dashed line}=[-, dashed]
 \usepackage{braket}
\usepackage{physics}
\usepackage{luatex85,qcircuit}
\usepackage{blkarray}

\theoremstyle{plain}
\newtheorem{axiom}{Axiom}
\newtheorem{theorem}{Theorem}

\newtheorem{lemma}{Lemma}

\theoremstyle{definition}
\newtheorem{definition}{Definition}

\newtheorem{problem}{Problem}

\newcommand{\hilbertspace}{\mathcal{H}}
\newcommand{\bigO}{\mathcal{O}}
\newcommand{\lagrangian}{\mathcal{L}}

\newcommand{\realnumber}{\mathbb{R}}

\newcommand{\integer}{\mathbb{Z}}

\newcommand{\identity}{\mathds{1}}

\newcommand{\ii}{\textup{i}}

\newcommand{\probability}{\mathbb{P}}

\newcommand{\cnot}{\textup{\textsc{cnot}}}
\newcommand{\hdm}{\textup{\textsc{h}}}

\newcommand{\negate}{\textup{\textsc{not}}}

\newcommand{\maxcut}{\textup{\textsc{MaxCut}}}
\newcommand{\sat}{\textup{\textsc{sat}}}

\let\cclass\textup
\let\P\relax
\newcommand{\P}{\cclass{P}}
\newcommand{\PP}{\cclass{PP}}

\newcommand{\sharpP}{\cclass{\#P}}

\newcommand{\BQP}{\cclass{BQP}}

\newcommand{\PSPACE}{\cclass{PSPACE}}

\newcommand{\T}{\intercal}

\newcommand\vartextvisiblespace[1][.5em]{\makebox[#1]{\kern.07em
    \vrule height.3ex
    \hrulefill
    \vrule height.3ex
    \kern.07em
  }}

\newcommand{\zpartition}{\mathcal{Z}}
\newcommand{\llaplacian}{\mathfrak{L}}
\newcommand{\dlagrangian}{\mathcal{L}}
\newcommand{\eaction}{\mathcal{S}_{\textup{E}}}
\newcommand{\action}{\mathcal{S}}
\newcommand{\hhat}{\hat{H}}
\newcommand{\xhat}{\hat{x}}
\newcommand{\phat}{\hat{p}}

\newcommand{\vbx}{\vb{x}}

\newcommand{\vbq}{\vb{q}}

\newcommand{\hatsigma}{\hat{\sigma}}

\newcommand{\sx}{\hat{\sigma}_x}
\newcommand{\sy}{\hat{\sigma}_y}
\newcommand{\sz}{\hat{\sigma}_z}

\newcommand{\px}{\hat{X}}
\newcommand{\py}{\hat{Y}}
\newcommand{\pz}{\hat{Z}}
\newcommand{\pI}{\hat{I}}
\newcommand{\schrodinger}{\textup{Schr\"{o}dinger }}

\newcommand{\deltat}{\Delta t}
\newcommand{\deltatau}{\Delta \tau}

\usepackage{stmaryrd}
 \renewcommand{\llaplacian}{\hat{\mathfrak{L}}}

\newcommand{\U}{\hat{U}}
\newcommand{\oracle}{\hat{O}}
\newcommand{\D}{\mathcal{D}}

\newcommand{\intinf}{\int_{-\infty}^{\infty}}

\makeatletter
\def\l@subsubsection#1#2{}
\makeatother

\begin{document}
\title{On Lagrangian Formalism of Quantum Computation}
\author{Jue Xu}
\email{juexu@cs.umd.edu}
\date{\today}
\begin{abstract}
We reformulate quantum computation in terms of Lagrangian (sum-over-path) formalism, in contrast to the widely used Hamiltonian (unitary gate) formulation.
	We exemplify this formalism with some widely-studied models, 
	including the standard quantum circuit model, quantum optimization heuristics, and quantum random walk.
	The meanings of Lagrangian (action) are interpreted in various contexts of quantum computation, such as complexity.
	Furthermore, an analog quantum simulation scheme is suggested where the Lagrangian serves as the starting point and the sum-over-path method is applied.
\end{abstract}

\maketitle
\tableofcontents

\section{Introduction}
Quantum algorithms, especially the ones modeled by quantum circuits \cite{nielsenQuantumComputationQuantum2010}, 
have been widely written in the language of the \schrodinger picture, 
where the computation process is a unitary evolution of quantum superposition states and the unitary evolution is commonly decomposed into a set of local unitary gates.
This circuit formulation is essentially successful 
because it is a well-developed framework
with plenty of powerful tools inherited from classical computation,
including universal computation, error correction, and provable algorithmic speedup.
However, the standard circuit model is not the mere choice for quantum computation. 
Many insightful and powerful models, like adiabatic quantum computation \cite{farhiQuantumComputationAdiabatic2000}, quantum random walks \cite{childsQuantumInformationProcessing2004} \cite{ambainisOnedimensionalQuantumWalks2001}, 
topological quantum computation \cite{kitaevFaulttolerantQuantumComputation2003}, and measurement-based quantum computation \cite{briegelMeasurementbasedQuantumComputation2009}, etc,
bring their own advantages.
For example, the Heisenberg representation of quantum computers describing the evolution of operators rather than states has proven extremely important in understanding quantum error correction \cite{gottesmanHeisenbergRepresentationQuantum1998}.
So, it is always worthwhile to seek new formulation of quantum computation.

Lagrangian formalism is a natural candidate.
Lagrangian formalism of physics is known to be equivalent to the canonical (Hamiltonian) formalism,
but two formalisms have their own pros and cons.
For example, Hamiltonian formalism explains unitarity of evolution and spectrum quite well.
On the other hand, it is easier to discuss symmetry and (relativistic) Lorentz, gauge invariance with Lagrangian formalism
such that Lagrangian (action) is preferable to Hamiltonian as the starting point of advanced physics like quantum field theory.
Lagrangian formalism of quantum mechanism is also known as \nameref{thm:path_integral} formalism,
which is a remarkable generalization of \nameref{thm:least_action} in classical mechanics.
Although Lagrangian formalism of quantum computation has not been proposed systematicly, 
the idea of sum-over-path (path integral) has been widely applied in different areas of quantum computation,
such as an alternative formulation of quantum circuits \cite{penneyQuantumCircuitDynamics2017}
and proofs of quantum complexity bounds \cite{bernsteinQuantumComplexityTheory1997}\cite{dawsonQuantumComputingPolynomial2004}.
The view of path integral can help explain the magic of quantum computation in an intuitive and visual way:
sophisticated quantum algorithms explore all possible paths simultaneously
and undesired paths are canceled by interference.

In this article, we are not only interested in the Lagrangian formalism for its use as an alternative description of quantum computation, 
but rather for the novel perspective that it provides.
Intuitively, action measures the runtime of quantum computation,
while Lagrangian quantifies its growth rate.
On the other hand, action itself can be regarded as an objective of intractable optimization problems.
Moreover, quantum simulation algorithms can be designed with Lagrangian instead of Hamiltonian for certain system e.g. QED,
where Lagrangian is originally the starting point of the system.
As the goal of the paper is primarily to demonstrate the potential of a new framework, 
we will not put much attention to technical issues of algorithms.
One motivation of our formulation is that it allows the tools of quantum physics to be applied to understand quantum computation.

The structure of the paper is as follows.
In the remainder of this section, we will clarify some notations and conventions,
and followed by a quick review on Lagrangian formalism.
\cref{sec:review_circuit} reformulates the circuit model with sum-over-path techique 
and examines its applications to simple textbook problems.
In \autoref{sec:qaoa}, we discuss optimality (complexity) of protocols of quantum heuristics in terms of action.
\cref{sec:walk} illustrates continuous-time quantum random walk as quantum propagator on graphs, and the relation between discrete-time quantum random walk and Dirac equation by the sum-over-path argument. 
\cref{sec:simulation} suggests a quantum simulation scheme based on Lagrangians and sum-over-path on lattice.
At last, in \autoref{sec:discussion}, we make remarks about potentials of this framework that remains to explore.
We assume throughout that the reader is familiar with the basics of quantum computing such as Pauli gates, universal gate set and textbook quantum algorithms at the level of \cite{nielsenQuantumComputationQuantum2010}. 
Nevertheless, this paper is still highly cross-field.
Therefore, we include considerable introductory material in the appendices.
A pedagogical review on path integral formalism for different kinds of physical systems is presented in \autoref{sec:path_integral}.
Basic concepts and definitions of computation complexity theory mentioned in the main text are briefly introduced in \autoref{sec:complexity_theory}. 

\subsection{Notations and conventions}
Since this paper interleaves physics and computer science deeply
such that one quantity/operator might have different notations in different contexts, 
we give several clarifications about our conventions:
\begin{itemize}[leftmargin=*]
    \item 
The hats on $\hhat$ and $\U$ emphasize that the unitary gates and quantum Hamiltonians are operators (matrices)
	in contrast to the classical Hamiltonian $H$ and Lagrangian $\dlagrangian$ that are scalar functions.
	For the sake of simplicity, we usually drop the identity term in a quantum Hamiltonian, 
	because it only contributes a global phase to states.
	We denote this equivalence as $\hhat\simeq C\cdot\identity + \hhat$ 
where $C$ is a configuration independent constant.
	Similarly, constant terms in an action $\action$ can be omitted.
\item
	Rather than Pauli matrices commonly denoted as $\qty{\hat{\sigma}_0\equiv\pI,\hatsigma_1\equiv\sx,\hatsigma_2\equiv\sy,\hatsigma_3\equiv\sz}$ in physics,
	we adopt the notation $\qty{\pI,\px,\py,\pz}$ in this paper to save room for indices.
Unless otherwise stated, the subscript of the Pauli gates is reserved for the index of the qubit and the superscript for time step.
That is, $\pz_j^{(l)}$ denotes the Pauli $Z$ operator acting on the $j$-th qubit at $l$ time step.

    \item 
    In physics, the classical state of a two-level (e.g. Ising spin) system is commonly described by a scalar $\sigma\in\qty{1,-1}$, 
while one qubit state $\ket{q}$ is described by the eigenstates (vectors) of $\pz$, i.e.,
    $\qty{\ket{\uparrow}\equiv \ket{0}, \ket{\downarrow}\equiv \ket{1}}$.
    The classical state $\sigma$ can be regarded as the eigenvalue of a qubit state, namely,
    $\pz\ket{q}=\sigma\ket{q}=e^{\ii \pi q}\ket{q}$.
For an $n$-qubit system, we denote the computational basis of $2^n$ states as $\ket{z}\equiv \ket{\vbq}\equiv\ket{q_1}\dots\ket{q_n}$ 
    where $z\in [2^n]\equiv \qty{0,1,\dots,2^n-1}$ and $\vbq\equiv q_1 \dots q_n$ is the binary representation of $z$.

\end{itemize}

\subsection{Lagrangian formalism: from Euler-Lagrange equation to path integral}
In optics, Fermat's principle states that the path taken by a ray between two given points is the path that can be traveled in the least (extremum) time. 
A similar argument, \emph{principle of least action}, was developed in classical mechanics:
\begin{axiom}[Principle of least action]\label{thm:least_action}
    The actual path $q(t)$ taken by a classical system is the path that 
	yields an extremum of its action \(\action\).
	So, this principle is also called principle of stationary action.
The action of the dynamics is the integral of Lagrangian over time
	\begin{equation}
		\action[q(t)]:=\int_{t_I}^{t_F}\dlagrangian(q(t),\dot{q}(t);t)\dd{t}
		\label{eq:action}
	\end{equation}
	where $\lagrangian(q,\dot{q})$ is the Lagrangian in terms of generalized coordinate $q$ and velocity $\dot{q}$ at certain time $t$. 
\end{axiom}
The notion $\action[\cdot]$ reminds that action is a functional that takes a function (path) $q(t)$ as input.
By varying the action, one have the equation of motion (Eq.\ref{eq:euler_lagrange}) called Euler-Lagrange equation.
This Lagrangian formalism was extended by Dirac \cite{diracAnalogyClassicalQuantum1945} and Feynman \cite{feynmanQuantumMechanicsPath2010} to explain quantum mechanics. 
\begin{axiom}[Path integral]\label{thm:path_integral}
    The amplitude (probability) of a quantum system evolving from $\ket{q_I}$ to $\ket{q_F}$ in a time interval can be evaluated by (functional) integrating over all possible paths with fixed initial and final position 
    \begin{equation}
		\mel{q_F}{e^{-\ii t\hhat/\hbar}}{q_I} =
        \int_{q(t_I)= q_F}^{q(t_F)=q_I} \D q \; e^{\ii \action[q]/\hbar}
    \end{equation}
	where the action defined in classical mechanics as \cref{eq:action}.
\end{axiom}
The Larangian (path integral) formalism of quantum mechanics is proved to be equivalent to the well-known \schrodinger equation \cite[Chp4]{feynmanQuantumMechanicsPath2010} 
\begin{equation}
    \ii\hbar \dv{t} \ket{\psi(t)} = \hhat(t) \ket{\psi(t)}
\label{eq:evolution}
\end{equation}
which is a differential equation determining the evolution of quantum state.
In the classical limit (Planck's constant $\hbar\to 0$), \nameref{thm:path_integral} reduces to \nameref{thm:least_action}
because only the paths around the stationary point of the action contribute 
(the other paths' contributions frequently oscillate and cancel out).
We have included a summary of the path integral formalism for various kinds of systems in \cref{sec:path_integral}.

\section{A Reformulation of Quantum Circuit Model}\label{sec:review_circuit}
The quantum circuit model serves as the standard model of quantum computation
because it is powerful and well-established from the viewpoint of theoretic computer science. 
However, decomposing the computational process into elementary gates makes the design of quantum algorithms a task of computer science or applied mathematics, losing its intrinsic connection to physics.
Although we don't expect that an alternative model of quantum computation would provide more computational power than the standard one,
a new formulation would provide a different view.
This point of view would inspire a new paradigm of quantum algorithm
and it would reveal a connection between quantum physics and computational complexity.
In this section, we will reformulate the standard quantum circuit model and revisit two keystone algorithms proposed in the circuit (oracle) model.

\subsection{Sum-over-path formulation of quantum circuits}
\begin{figure}[!ht]
    \centering
    \begin{subfigure}{0.49\textwidth}
	\centering
	\begin{equation*}
		\Qcircuit @C=1.1em @R=1em {
			&&&&\mbox{Diffusion operator}&&& \\
		\lstick{\ket{0}}&\gate{\hdm}&\multigate{2}{\oracle_w}&\gate{\hdm}&\multigate{2}{\U_A}&\gate{\hdm}&\multigate{2}{\oracle_w}&\qw&\cdots \\
						&\vdots  &\ghost{\oracle_w}       &\vdots  &\ghost{\U_A}       &\vdots  &\ghost{\oracle_w}       &\qw&\cdots\\
		\lstick{\ket{0}}&\gate{\hdm}&\ghost{\oracle_w}       &\gate{\hdm}&\ghost{\U_A}       &\gate{\hdm}&\ghost{\oracle_w}       &\qw&\cdots
			\gategroup{2}{4}{4}{6}{.7em}{--} 
}
	\end{equation*}
\caption{The standard circuit of Grover's algorithm.}
    \end{subfigure}
\begin{subfigure}{0.49\textwidth}
    \centering
    \ctikzfig{grover}
    \caption{Distribution of amplitudes of Grover's algorithm with $w=1$, $N=4$. The dashed line indicates the average of all amplitudes.}
    \end{subfigure}
	\caption{Two views of Grover's algorithm.}
	\label{fig:grover}
\end{figure}
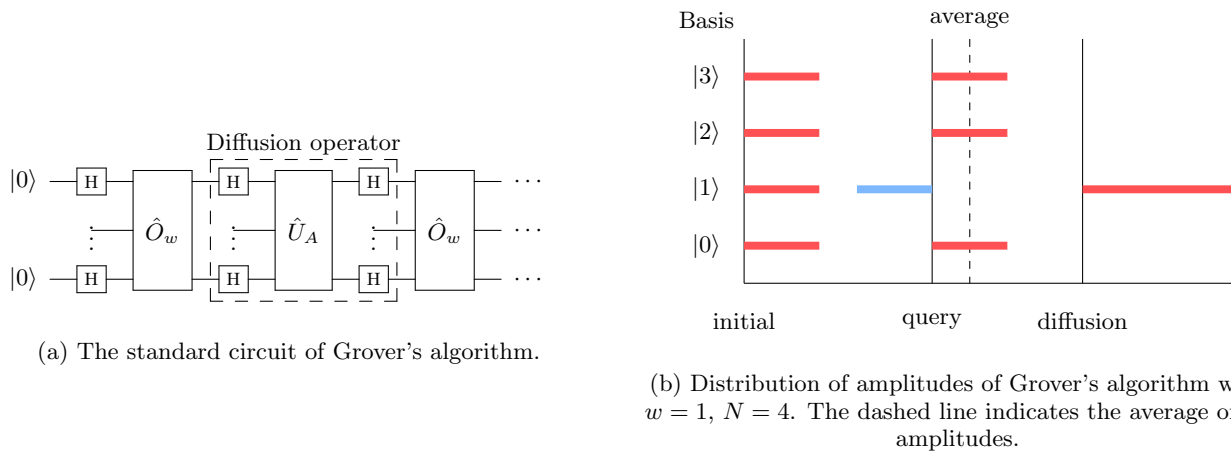

There are papers 
\cite{dawsonQuantumComputingPolynomial2004} 
\cite{rudiak-gouldSumoverhistoriesFormulationQuantum2006} 
and
\cite{penneyQuantumCircuitDynamics2017}
formulating quantum circuits based on sum-over-path technique.
The basic idea of our paper is similar to theirs, 
but details are different. 
For example, in our formulation the unitaries are applied to all computational basis instead of local qubits as in these papers.
This change makes the interference of paths presented in an intuitive and visual manner
(e.g. \cref{fig:deutsch} and \cref{fig:grover}).
In addition, in our paper, the sum-over-path formulation is not only applied to the circuit model, but also other universal models such as adiabatic quantum computation and quantum random walks.
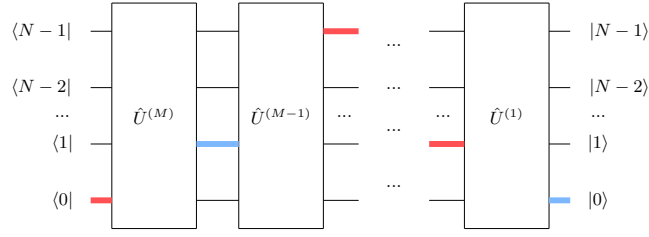
\begin{figure}[!ht]
    \centering
    \scalebox{0.75}{\begin{tikzpicture}
	\begin{pgfonlayer}{nodelayer}
		\node [style=none] (0) at (-3.5, 4) {};
		\node [style=none] (2) at (-0.5, 4) {};
		\node [style=none] (3) at (-0.5, -4) {};
		\node [style=none] (4) at (-3.5, -4) {};
		\node [style=none] (5) at (-2, 0) {$\hat{U}^{(M-1)}$};
		\node [style=none] (23) at (-0.5, -1) {};
		\node [style=none] (24) at (-0.5, 1) {};
		\node [style=none] (27) at (-3.5, -1) {};
		\node [style=none] (28) at (-3.5, 1) {};
		\node [style=none] (41) at (4.5, 4) {};
		\node [style=none] (42) at (7.5, 4) {};
		\node [style=none] (43) at (7.5, -4) {};
		\node [style=none] (44) at (4.5, -4) {};
		\node [style=none] (45) at (6, 0) {$\hat{U}^{(1)}$};
		\node [style=none] (51) at (4.5, -1) {};
		\node [style=none] (52) at (4.5, 1) {};
		\node [style=none] (53) at (-8, 4) {};
		\node [style=none] (54) at (-5, 4) {};
		\node [style=none] (55) at (-5, -4) {};
		\node [style=none] (56) at (-8, -4) {};
		\node [style=none] (57) at (-6.5, 0) {$\hat{U}^{(M)}$};
		\node [style=none] (59) at (-5, -1) {};
		\node [style=none] (60) at (-5, 1) {};
		\node [style=none] (65) at (-5, 3) {};
		\node [style=none] (66) at (-3.5, 3) {};
		\node [style=none] (67) at (-5, -3) {};
		\node [style=none] (68) at (-3.5, -3) {};
		\node [style=none] (71) at (-0.5, 3) {};
		\node [style=none] (76) at (4.5, 3) {};
		\node [style=none] (78) at (4.5, -3) {};
		\node [style=none] (79) at (-0.5, -3) {};
		\node [style=none] (81) at (-8.75, 3) {};
		\node [style=none] (82) at (-8, 3) {};
		\node [style=none] (93) at (-8.75, 1) {};
		\node [style=none] (94) at (-8, 1) {};
		\node [style=none] (95) at (-8.75, -1) {};
		\node [style=none] (96) at (-8, -1) {};
		\node [style=none] (99) at (-8.75, -3) {};
		\node [style=none] (100) at (-8, -3) {};
		\node [style=none] (102) at (-10.5, 3) {$\bra{N-1}$};
		\node [style=none] (103) at (-10.5, 1) {$\bra{N-2}$};
		\node [style=none] (104) at (-9.75, -1) {$\bra{1}$};
		\node [style=none] (106) at (-9.75, -3) {$\bra{0}$};
		\node [style=none] (109) at (-9.75, 0) {...};
		\node [style=none] (110) at (7.5, -1) {};
		\node [style=none] (111) at (7.5, 1) {};
		\node [style=none] (115) at (8.25, -1) {};
		\node [style=none] (116) at (8.25, 1) {};
		\node [style=none] (117) at (7.5, 3) {};
		\node [style=none] (118) at (8.25, 3) {};
		\node [style=none] (119) at (7.5, -3) {};
		\node [style=none] (120) at (8.25, -3) {};
		\node [style=none] (125) at (0.75, 3) {};
		\node [style=none] (126) at (0.75, 1) {};
		\node [style=none] (127) at (0.75, -1) {};
		\node [style=none] (129) at (0.75, -3) {};
		\node [style=none] (130) at (3.25, 3) {};
		\node [style=none] (131) at (3.25, 1) {};
		\node [style=none] (132) at (3.25, -1) {};
		\node [style=none] (134) at (3.25, -3) {};
		\node [style=none] (135) at (2, 2.5) {...};
		\node [style=none] (136) at (2, 1) {...};
		\node [style=none] (137) at (2, -0.5) {...};
		\node [style=none] (139) at (2, -2.5) {...};
		\node [style=none] (155) at (0.25, 0) {...};
		\node [style=none] (156) at (3.75, 0) {...};
		\node [style=none] (157) at (10, 3) {$\ket{N-1}$};
		\node [style=none] (158) at (10, 1) {$\ket{N-2}$};
		\node [style=none] (159) at (9.25, -1) {$\ket{1}$};
		\node [style=none] (160) at (9.25, -3) {$\ket{0}$};
		\node [style=none] (161) at (9.25, 0) {...};
	\end{pgfonlayer}
	\begin{pgfonlayer}{edgelayer}
		\draw (0.center) to (4.center);
		\draw (4.center) to (3.center);
		\draw (0.center) to (2.center);
		\draw (2.center) to (3.center);
		\draw (41.center) to (44.center);
		\draw (44.center) to (43.center);
		\draw (41.center) to (42.center);
		\draw (42.center) to (43.center);
		\draw (53.center) to (56.center);
		\draw (56.center) to (55.center);
		\draw (53.center) to (54.center);
		\draw (54.center) to (55.center);
		\draw (60.center) to (28.center);
		\draw (65.center) to (66.center);
		\draw (81.center) to (82.center);
		\draw (93.center) to (94.center);
		\draw (95.center) to (96.center);
		\draw (111.center) to (116.center);
		\draw (110.center) to (115.center);
		\draw (117.center) to (118.center);
		\draw (24.center) to (126.center);
		\draw (23.center) to (127.center);
		\draw (79.center) to (129.center);
		\draw (130.center) to (76.center);
		\draw (134.center) to (78.center);
		\draw [style=bold red edge] (99.center) to (100.center);
		\draw [style=bold red edge] (71.center) to (125.center);
		\draw [style=bold red edge] (132.center) to (51.center);
		\draw [style=bold blue] (119.center) to (120.center);
		\draw [style=bold blue] (59.center) to (27.center);
		\draw (131.center) to (52.center);
		\draw (67.center) to (68.center);
	\end{pgfonlayer}
\end{tikzpicture}}
\caption{Sum-over-path formulation of $M$-depth quantum circuits: Each horizontal line represents one computational basis $\ket{z}$. Two colors indicate the (positive, negative) sign of amplitudes.}
	\label{fig:path_circuit}
\end{figure}
The physical meaning of action and Lagrangian in quantum computation will be discussed extensively in the subsequent sections.

In the standard quantum circuit model, the whole circuit $\U$ of an algorithm is a sequence of elementary, (local) unitary operators as shown in \cref{fig:path_circuit}
\begin{equation}
    \U = \U^{(M)}\U^{(M-1)}\cdots\U^{(2)}\U^{(1)}
	\equiv\prod_{l=1}^M \U^{(l)}.
\end{equation}
For simplicity, we can assume $\forall l,\U^{(l)}\in \qty{\U_{\hdm},\U_{\textsc{Toffoli}}}$ which is known as a universal gate set \cite{shiBothToffoliControlledNOT2002}.
The Toffoli gate is famous as a universal, reversible classical gate,
while the Hadamard gate is responsible for the quantum interference in quantum computation.
In the path view of quantum circuits,
the Hadamard gate redistributes signed amplitudes from one computational basis to two, while the Toffoli gate only transports the amplitude from one to another, e.g.,
\begin{equation}
    \mel{q_1'q_2'q_3'}{\U_{\textup{Toffoli}}}{q_1q_2q_3}
    =
    \delta_{q_1',q_1} \delta_{q_2',q_2} \delta_{q_3',q_1q_2\oplus q_3}.
	\label{eq:toffoli}
\end{equation}
In other words, the classical gates, such as CNOT and Toffoli gates, set contrainsts on allowed paths.

\subsubsection{Actions of quantum gates and circuit}
``Action" here has two-fold meanings, one is ``behavior" and the other is ``integral of Lagrangian over time".
As a warmup, look at the simple but indispensable quantum gate, i.e., Hadamard gate $\U_{\hdm}$,
which can be interpreted as quantum Fourier transform (QFT) over $\integer_2$:
\begin{equation}
    \U_{\hdm} \equiv \text{QFT}(\integer_2) =
    \frac{1}{\sqrt{2}}
    \pmqty{e^{\ii\pi 0\cdot 0} & e^{\ii\pi 0\cdot 1}\\e^{\ii\pi 1\cdot 0}&e^{\ii\pi 1\cdot 1} }
\end{equation}
The matrix element of $\U_{\hdm}$ (propagator) can be written as the exponential of an integral along ``one path"
\begin{align*}
    \mel{q'}{\U_{\hdm}}{q} 
&= \frac{1}{\sqrt{2}} e^{\ii \pi (q'\cdot q)}
    =\frac{1}{\sqrt{2}} \exp(\ii \int_0^{\pi} q'\cdot q \dd{t})
\end{align*}
where $q'\cdot q$ is a time-independent ``Lagrangian" and evolution time is $\pi$.
This form can be directly generalized to the tensor product of $n$ Hadamard gate, i.e., the QFT over the number field $(\integer_2)^n$
\begin{align}
    \mel{\vbq' }{\U_\hdm^{\otimes n}}{\vbq }
&= \frac{1}{\sqrt{2^n}} e^{\ii \pi \sum_{j}^n q_j\cdot q_j'}
,\;
	\vbq\in \qty{0,1}^n 
\label{eq:qft_z2n}
\end{align}
Another important example of QFT is the one over the field $\integer_{2^n}$ 
\begin{equation}
	\mel{z'}{\text{QFT}(\integer_{2^n})}{z} 
	= \frac{1}{\sqrt{2^n}} e^{\ii 2\pi z'z/2^n}
	,\;
	z \in [2^n]
	\label{eq:qft_zN}
\end{equation}
which corresponds to the classical discrete (fast) fourier transform.

Now, we can evaluate the matrix element of the whole unitary $\U$ (circuit) by inserting $M-1$ sets of complete basis states (resolution of identity)
$\sum_{\vbq}\op{\vbq}{\vbq}=\pI^{\otimes n}$
(the trick widely used in \nameref{sec:path_integral})
\begin{align}
    \mel{\vbq^{(M)}}{\U}{\vbq^{(0)}}
=& 
	\sum_{ \qty{\vec{\vbq}} } \prod_l^{M} \mel{\vbq^{(l)}}{\U^{(l)}}{\vbq^{(l-1)}}
\end{align}
where $\vec{\vbq}=(\vbq^{(1)},\dots,\vbq^{(M-2)},\vbq^{(M-1)})$ describes a discrete path 
and each ``coordinate" $\vbq^{(l)}\in \qty{0,1}^n$ is a binary string.
The acronym $\sum_{\qty{\vec{\vbq}}} := \sum_{\vbq^{(M-1)}}\sum_{\vbq^{(M-2)}}\dots\sum_{\vbq^{(1)}}$ stands for the sum over all paths, that is, the discrete version of path integral measure $\int \D x$.
The sum-over-path form of the circuit now reads
\begin{align}
    \mel{\psi_F}{\U}{\psi_I}
= \sum_{\text{allowed paths}} e^{\ii \action[\vec{\vbq};\U]}
	\label{eq:circuit_propagator}
\end{align}
where ``allowed paths" constraint is the result of all classical gates $\U_{\textsc{Toffoli}}$ and the ``action" of the circuit $\U$ is contributed by quantum gates $\U_{\hdm}$
\begin{equation}
    \action[\vec{\vbq};\U]
	:= \sum_{l=1}^M \dlagrangian^{(l)} \qty(\vbq^{(l-1)},\vbq^{(l)})
	,\;
	\dlagrangian=(q'\cdot q) \pi .
\end{equation}
In contrast to that quantum circuits (unitary) and quantum Hamiltonian (hermitian) are matrices,
Lagrangian is a scalar function $\dlagrangian: \qty{0,1}^n\times \qty{0,1}^n \to \qty{0,\pi}$.
The meaning of this discrete action is not manifest when we reformulate quantum circuits,
but we will interpret it as complexity in \autoref{sec:qaoa}.

\subsubsection{Toy example of interference of paths: QFT and Deutsch's algorithm}
Quantum Fourier transform (QFT) is one cornerstone of quantum algorithms,
including Shor's algorithm \cite{shorPolynomialTimeAlgorithmsPrime1997} (finding the prime factors of an integer) which exhibts the exponential separation between classical and quantum computation.
The speedup by QFT comes from quantum interference of paths and group structures of the problems studied.
To see the effect of quantum interference,
the simplest example is two sequential Hadamard gates equal to identity operator:
orthogonality of basis corresponds to interferenece of two paths
\begin{align*}
    \mel{q'}{\U_{\hdm}\U_{\hdm}}{q} 
    & = \sum_{q''\in \qty{0,1}} \mel{q'}{\U_{\hdm}\op{q''}{q''}\U_{\hdm}}{q} 
\\
& = \sum_{\sigma\in\qty{+,-}} \ip{q'}{\sigma} \ip{\sigma}{q}
= \frac{1}{2} \sum_{q''\in \qty{0,1}} e^{\ii\pi (q'-q)\cdot q''}
    \\
    & = \frac{1}{2} \qty( e^{\ii\pi (q'-q)} + 1)
    = \delta_{q',q}
	= \ip{q'}{q} 
\end{align*}
where $\ket{\sigma}:=\U_{\hdm}\ket{q}$ is another set of complete basis states,
namely, the QFT($\integer_2$) of computational basis $\ket{q}$ (eigenstate of ``momentum'' operator $\px$).
Inserting a complete basis of momentum is also commonly used in path integral formalism. 

The algorithms based on QFT are widely studied in so-called query (black-box/oracle) model
where the input problem is wrapped into a black-box.
The complexity of an algorithm is measured by the number of queries required to determine certain property of the input in the worst case
and the complexity of the problem is the complexity of the optimal algorithm.
A quantum algorithm in query model is viewed as the alternating oracles $\oracle$ and arbitrary unitaries $\U$
\begin{equation}
    \ket{\psi_F}= \U^{(M)} \oracle \U^{(M-1)} \dots \U^{(2)} \oracle \U^{(1)} \ket{\psi_I}
\end{equation}
where the initial state $\ket{\psi_I}$ usually takes $\ket{0^{\otimes n}}$.
Measuring the final state $\ket{\psi_F}$ gives the answer to the problem.

The simplest quantum algorithm might be Deutsch's algorithm 
that determines certain property (balanced or constant) of a black-box Boolean function $f$ by utilizing QFT over group $\integer_2$.
The bit-flip oracle $\hat{O}_f$ encoding $f$ acts as 
\begin{equation}
	\hat{O}_{f} \ket{z,q} = \ket{z,q\oplus f(z)}
	,\; z \in [N],\, q\in \qty{0,1}
	\label{eq:bit_flip_oracle}
\end{equation}
where $f(z):[N]\to \qty{0,1}$ is the black-box Boolean function.
This is simply the quantum version of the classical reversible oracle mapping $(z,q)\mapsto (z, q\oplus f(z))$.
To see the behavior of the algorithm straightforwardly, 
we instantiate the oracle $\oracle$ by a $\cnot$ gate which means the black-box function $f$ is $\textsc{not}$ (bit-flip) gate, which is a balanced function.
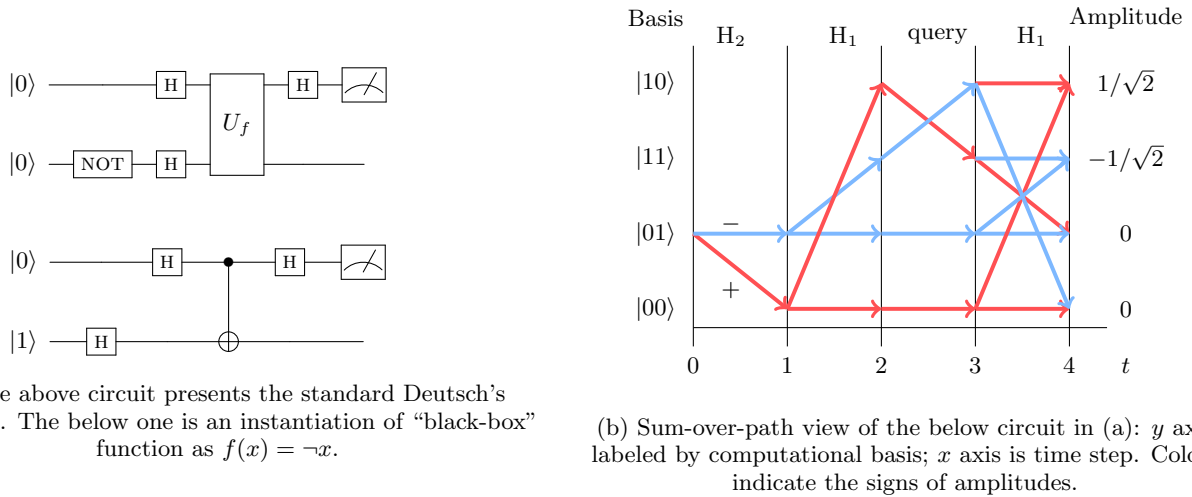
\begin{figure}[!ht]
    \centering
    \begin{subfigure}{0.49\textwidth}
    \centering
    \begin{equation*}
        \Qcircuit @C=1em @R=2em {
        \lstick{\ket{0}}&\qw &\gate{\hdm} & \multigate{1}{U_f} & \gate{\hdm}&\meter \\
        \lstick{\ket{0}}&\gate{\negate} & \gate{\hdm} & \ghost{U_f}& \qw&\qw}
    \end{equation*}
	\vspace{.5cm}
    \begin{equation*}
    \Qcircuit @C=1.5em @R=2em {
    \lstick{\ket{0}} & \qw & \gate{\hdm} & \ctrl{1} & \gate{\hdm}&\meter \\
    \lstick{\ket{1}} &\gate{\hdm} & \qw & \targ & \qw &\qw
    }
    \end{equation*}
    \caption{The above circuit presents the standard Deutsch's algorithm. The below one is an instantiation of ``black-box" function as $f(x)=\neg x$.}
    \end{subfigure}
	\hfill
    \begin{subfigure}{0.49\textwidth}
    \centering
    \ctikzfig{circuit_path}
    \caption{Sum-over-path view of the below circuit in (a): $y$ axis labeled by computational basis; $x$ axis is  time step. Colors indicate the signs of amplitudes.}
    \end{subfigure}
	\caption{Circuit view and path view of Deutsch's algorithm.}
	\label{fig:deutsch}
\end{figure}
In \cref{fig:deutsch}, it is clear to see quantum parallelism together with interference of paths enables the speedup (where classical computation requires two queries).
Conversely, if the black-box function is a constant function (e.g. identity),
we will see a different pair of paths canceled giving a different answer.

\subsubsection{Grover search: diffusion over basis and flip the target}
Another cornerstone quantum algorithm, Grover's algorithm \cite{groverQuantumMechanicsHelps1997} exhibits the quadratic speedup (separation) over classical computation in the unstructured search problem.
In Grover's algorithm, the oracle is phase-flip oracle 
\begin{equation}
    \hat{O}_f : \ket{z} \mapsto (-1)^{f(z)} \ket{z}
    = e^{\ii \pi f(z)} \ket{z}
    ,\, f(z)\in \qty{0,1}
\end{equation}
which is equivalent to bit-flip oracle \cref{eq:bit_flip_oracle} up to Hadamard conjugation 
(phase kickback trick: $\U_{\hdm}\pz\U_{\hdm}\equiv\px$).
The Grover oracle flips the sign of the marked basis $\oracle_w \ket{z} = (-1)^{\delta_{z,w}}\ket{z}$, more specifically
\begin{equation}
    \oracle_w
    := \pI^{\otimes n} - 2 \op{w}{w}
\label{eq:grover_oracle_unitary}
\end{equation}
where $\ket{w}$ is the unknown marked state (one computational basis). 
The other ingredient of Grover's algorithm is the diffusion operator 
\begin{equation}
    \U_D:=\U_\hdm^{\otimes n} \U_A \U_\hdm^{\otimes n}
	=2\op{s}{s}-\pI^{\otimes n}
	\label{eq:grover_diffusion_unitary}
\end{equation}
with $\U_A:= 2 \op{0^{\otimes n}}{0^{\otimes n}} -\pI^{\otimes n}$
and the uniform superposition $\ket{s}:=\frac{1}{\sqrt{N}}\sum_z^N \ket{z}$.
We assume $N\equiv 2^n$ is the number of the items and $n$ is the number of qubits used to represent all items.
Focus on single iteration,
the propagator reads
\begin{align*}
    \mel{z^{(l)}}{ \U_{D} \oracle_w }{z^{(l-1)}}
=& 
    (-1)^{\delta(w,z^{(l-1)})} 
    \qty( 2/N - \delta(z^{(l)},z^{(l-1)}) )
	\label{eq:grover_lagrangian}
\end{align*}
Geometrically, every Grover iteration is two consecutive reflections/rotations $\U_D \oracle_w = e^{\ii \theta\py}$ in two-dimensional subspace $\qty{\ket{w},\ket{w^\perp}}$ spanned by the targe basis and the rest.
From the path view, the behavior of one Grover iteration is:
diffusion operator subtracts the amplitude of every basis from the double of the average of all amplitudes,
and the oracle flips the sign of the amplitude of the targe basis.
We will discuss more about complexity and optimality of Grover's search with respect to action in \autoref{sec:qaoa}
and we will also interpret it as quantum propagation on a complete graph with a delta potential in \autoref{sec:walk}. 

\subsection{Complexity implications}
In the circuit model, the computational complexity measures the total number of gates required to build the minimal circuit that generates the state encoding the answer.
In the query model, the goal of designing a algorithm is to find the optimal algorithm that approaches the lower bound of the query complexity of the problem.
By appealing to the idea of computational paths, 
Bernstein and Vazirani \cite{bernsteinQuantumComplexityTheory1997} proved the upper bounds on the power of Quantum Turing Machines, i.e.,
$\nameref{def:bqp}\subseteq \PSPACE$ and $\BQP\subseteq \P^{\sharpP}$.

With sum-over-path formulation, Dawson et al. \cite{dawsonQuantumComputingPolynomial2004} proved that
determining the output of a quantum computation (computing the propagator \cref{eq:circuit_propagator}) is equivalent to counting the number of solutions to certain sets of polynomial equations over the finite field $\integer_2$,
which is \nameref{def:sharp_p}-complete problem and the prototype of the so-called dynamical sign problem.
It implies that $\BQP\subseteq \P^{\sharpP}$ and a sharper result $\BQP\subseteq \PP$,
which means that classical simulation of universal quantum circuits is hard.
It is a typical paradigm of demonstrating quantum supremacy.
Roughly speaking, by relating a quantum computation task with a classical computation problem,
efficient classical simulation of quantum computation would imply the highly unlikely collapse of complexity hirerarchy.
Examples are
Boson sampling \cite{aaronsonComputationalComplexityLinear2011},
classical simulation of QAOA \cite{farhiQuantumSupremacyQuantum2016}, and
random circuit sampling \cite{boulandQuantumSupremacyComplexity2019}.
An application of Penney, Koh, and Spekkens's sum-over-path formalism \cite{penneyQuantumCircuitDynamics2017}
is an alternative proof \cite{kohComputingQuopitClifford2017} of
the Gottesman-Knill Theorem (``Heisenberg representation of quantum computer") \cite{gottesmanHeisenbergRepresentationQuantum1998}:
the outcome probabilities of Clifford circuits can be computed efficiently by classical computation.

\begin{table}[!ht]
\centering
\begin{tabular}{c|c|c}
    \hline
    \hline
    Models & Complexity measures  & Lagrangian \\  
    \hline
    Circuit/query &\#gate/depth/query & $\dlagrangian\in\qty{0,\pi}$ \\  
QAA/QAOA & spectrum gap/sum of angles  & (d+1) classical \\  
    Quantum walk & mixing time/spectrum & KG/Dirac \\  
\hline
\end{tabular}
\caption{Complexity measures in different models}
\end{table} 

As the thermodynamic quantity, entropy, plays an indispensable role in information theory and irreversible computation (Maxwell's demon),
several physical quantities are good candidates for complexity measures in quantum computation:
(1) The first option is quantum correlation, i.e., entanglement.
An intuition for this is that the sensitivity of a Boolean function measures how many bit flips affect the function value in the worst case, while correlation length can be thought of as a measure of the distance over which spins are affected by each other. 
More precisely, Ambainis's adversary method finds lower bounds of query complexities \cite{ambainisQuantumLowerBounds2002} by measuring entanglement increment during each query. 
Recently, Ji et al. \cite{jiMIPRE2020} show that allowing entanglement in quantum states shared by multiple parties greatly increases the power of quantum computation..
(2) The second choice is action/Lagrangian.
Close connections have been discovered between computational complexity and black hole in high-energy physics \cite{brownComplexityActionBlack2016}
\cite{brownSecondLawQuantum2018}, and in the context of the (AdS/CFT) correspondence \cite{boulandComputationalPseudorandomnessWormhole2019}.
In our Lagrangian formalism, it is natural to use
action as a complexity measure and Lagrangian as growth rate of complexity \cite{toffoliWhatLagrangianCounting2003}. 
However, it is nontrivial to carefully define action as a complexity measure to satisfy the properties of a complexity measure, such as positivity and monotonicity etc.
Once the connection between physical quantities (e.g. entanglement, action) and complexity is formally established,
we may bridge physical phase transitions and computational phase transitions naturally \cite{latorreAdiabaticQuantumComputation2004} \cite{deshpandeDynamicalPhaseTransitions2018},
where critical phenomena \cite{vidalEfficientClassicalSimulation2003} and scaling law \cite{orusUniversalityEntanglementQuantumcomputation2004} of computation can be studied with the tools from physics.
In the next section, we will see that action plays the role of complexity in some sense.

\section{Optimize Quantum Heuristics with Lagrangian}\label{sec:qaoa}
Quantum heuristic algorithms for optimization problems are a good platform for discussion on complexity in terms of Lagrangian and action.
Quantum adiabatic algorithm (QAA) is such a paradigm of quantum algorithm (actually univseral model \cite{aharonovAdiabaticQuantumComputation2007}) with the beautiful physical intuition,
originally proposed by Farhi et al. \cite{farhiQuantumComputationAdiabatic2000} for tackling satisfiability problem.
QAA finds the optimal solution of an optimization problem by evolving a quantum system slowly (adiabaticly) from the ground state of an easy-construct (trivial) Hamiltonian to the ground state of the final (problem) Hamiltonian that encodes the objective function.
In other words, the computation process is the evolution  governed by \schrodinger equation \cref{eq:evolution} and a time-dependent Hamiltonian that smoothly changes from $\hat{K}$ to $\hat{V}$, i.e.,
\begin{equation}
    \hhat_{}(t) = (1-\lambda(t)) \hat{K} + \lambda(t) \hat{V}
    \label{eq:adiabatic}
\end{equation}
where $\lambda(t)$ is a smooth function 
with fixed boundaries $\lambda(0)=0$ and $\lambda(T)=1$, assuming $T$ is the total evolution time.
Here, we use the kinetic energy notation $\hat{K}$ for the initial Hamiltonian and potential energy $\hat{V}$ for the final one instead of the common notation (i.e., $\hhat_B$ and $\hhat_C$).
This convention is at the moment not clear but we will find it suggestive.
A common choice of the initial (mixing) Hamiltonian (ought to contain non-zero off-diagonal entries)
is $\hat{K}=\sum_j \px_j$ whose ground state is the uniform superposition $\ket{s}$.
The final (problem) Hamiltonain $\hat{V}:=\sum_{\vbq} V(\vbq)\op{\vbq}{\vbq}$ is diagonal in computational basis (configuration) 
where the (classical) objective function $V(\vbq):\qty{0,1}^n\to \realnumber$ is evaluated.
A concrete example of this kind of problem is 3\nameref{def:sat} problem:
the (final) problem Hamiltonian $\hat{V}_{\Phi} = \sum_{c=1}^{m} \hat{V}_c$ is constructed as the sum of clause terms $\hat{V}_c$ corresponding to all $m$ clauses in the instance formula $\Phi$
where $\hat{V}_c$ acts on at most 3 qubits (identity on others).
In this setting, $V(\vbq)$ is the number of unsatisfied clauses with the assignment $\vbq$ and then the ground state of $\hat{V}$ naturally represents the optimal configuration (satisfying assignment).
According to the adiabatic theorem, the running (evolution) time $T$ of QAA is determined by the spectrum gap of $\hhat(t)$.
Unfortunately, for some systems of practical interest, the evolution time is exponential to the system size \cite{vandamHowPowerfulAdiabatic2001}.

To overcome the limitations of QAA and adapt to NISQ era (feasible qubit number, low circuit depth), 
Farhi et al. \cite{farhiQuantumApproximateOptimization2014} proposed Quantum Approximate Optimization Algorithm (QAOA) 
which allows an approximation ratio to the optimal solution.
In contrast to that QAA smoothly varies the Hamiltonian, 
QAOA interleaves two parametrized unitaries 
\begin{equation}
	\U_K(\beta):=e^{-\ii\beta\hat{K}}
	,\;
	\U_V(\gamma):=e^{-\ii\gamma\hat{V}}
\end{equation}
generated by two Hamiltonians ($\hat{K}$, $\hat{V}$) as defined in QAA and different angles $(\beta,\gamma)$.
For $2M$ angles $\pmb{\gamma} :=(\gamma^{(1)},\dots,\gamma^{(M)})$ and $\pmb{\beta} : = (\beta^{(1)},\dots ,\beta^{(M)})$,
we have the angle (parameter) dependent quantum state 
$\ket{\pmb{\gamma},\pmb{\beta}} = \prod^M_{l=1} \U_K(\beta^{(l)})  \U_V(\gamma^{(l)})\ket{\psi(0)}$
which gives the approximate optimal solution $\mel{\pmb{\gamma},\pmb{\beta}}{\hat{V}}{\pmb{\gamma},\pmb{\beta}}$ of the objective $V$.
The angles $\bm{\beta}$ and $\bm{\gamma}$ describe how long to apply each unitary (bang),
so the total runtime of QAOA is 
$T = \sum_{l=1}^M \qty(\gamma^{(l)} + \beta^{(l)})$.
It should be pointed out that the optimal angles for some problems can be found by classical computers efficiently.
In other words, the optimization of the QAOA protocol (runtime) is a classically tractable optimization problem.

\subsection{QAOA as Trotterization of QAA}
To implement QAA on conventional (standard circuit) quantum computers, the adiabatic quantum evolution (unitary) operator is supposed to be decomposed into a sequence of unitary operators each acting on a small number of qubits.
The common recipe is time slicing and Suzuki-Trotter product formula \cref{eq:product_formula} that are the key tricks in path integral formalism.
We have the discretized time evolution operator as
\begin{align*}
    e^{- \ii \int_0^T \hhat(t) \dd{t} }
&= \lim_{M\to \infty}
    \prod^M_{l=1} 
    \U_K\qty(\Delta \qty(1-\lambda\qty(l\Delta)) ) \U_V\qty(\Delta \lambda\qty(l\Delta) ).
\end{align*}
where $\Delta:=T/M$. 
The right hand side is a special case of QAOA with angles $\beta^{(l)}=\Delta (1-\lambda\qty(l\Delta)) $, $\gamma^{(l)}=\Delta \lambda(l\Delta)$, and large $M$.

An illustrative example is 2\nameref{def:sat} on a ring \cite{farhiQuantumComputationAdiabatic2000} \cite{wangQuantumApproximateOptimization2018} (a speical instance of \nameref{prm:maxcut}).
The objective (number of disagrees) to optimize is naturally written as the nearest-neighbor interaction of an Ising chain, 
so the two Hamiltonians of the QAA are
\begin{equation}
	\hat{K}_{\text{Ising}}=\sum_j^n \px_j
	,\quad
	\hat{V}_{\text{Ising}}=\sum_j^n \pz_j \pz_{j+1}
	\label{eq:ring_disagree}
\end{equation}
where the mixing (initial) Hamiltonian $\hat{K}_{\text{Ising}}$ is now interpreted as the external transverse field Ising energy.
From the point of view of graph theory, $\hat{K}=\sum_j^n \px_j$ is the adjacency matrix of hypercubes \cite{mooreQuantumWalksHypercube2001} corresponding to discrete kinetic energy, 
while $\hat{V}=\sum_j^n \pz_j \pz_{j+1}$ is a potential energy defined on hypercubes because it is diagonal with the computational basis (determined by a configuration of an Ising chain, a vertex of a hypercube).
\begin{figure}[!ht]
	\centering
	\scalebox{0.8}{\begin{tikzpicture}
	\begin{pgfonlayer}{nodelayer}
		\node [style=none] (20) at (-7.25, 6) {$q^{(l)}_{j}q^{(l)}_{j+1}$};
		\node [style=none] (25) at (1, 2) {};
		\node [style=none] (26) at (-5, 2) {};
		\node [style=none] (37) at (1, 10) {};
		\node [style=none] (38) at (-5, 10) {};
		\node [style=none] (41) at (-2, 2.25) {$q^{(l)}_{j}  q^{(l+1)}_{j}$};
		\node [style=none] (42) at (-2, 9) {};
		\node [style=none] (91) at (-7, 3) {$j$};
		\node [style=none] (92) at (-7, 9) {$j+1$};
		\node [style=none] (95) at (-6, 9) {};
		\node [style=none] (97) at (-6, 3) {};
		\node [style=none] (101) at (2, 9) {};
		\node [style=none] (103) at (2, 3) {};
		\node [style=bold vertex] (109) at (1, 3) {};
		\node [style=none] (127) at (-5, 10.75) {$\vbq$};
		\node [style=none] (133) at (3, 3) {$t$};
		\node [style=none] (138) at (1, 1.25) {$l+1$};
		\node [style=none] (140) at (-5, 1.25) {$l$};
		\node [style=bold vertex] (142) at (1, 9) {};
		\node [style=bold vertex] (143) at (-5, 9) {};
		\node [style=bold vertex] (144) at (-5, 3) {};
		\node [style=none] (145) at (-2, 3.75) {$\px_{j}$};
		\node [style=none] (146) at (-3.25, 6) {$\pz_{j}\pz_{j+1}$};
	\end{pgfonlayer}
	\begin{pgfonlayer}{edgelayer}
		\draw (38.center) to (26.center);
		\draw (37.center) to (25.center);
		\draw (95.center) to (101.center);
		\draw (97.center) to (103.center);
	\end{pgfonlayer}
\end{tikzpicture}}
	\caption{Correspondence between 1D transverse quantum Ising chain and (1+1)D classical Ising lattice.}
	\label{fig:plaquette}
\end{figure}
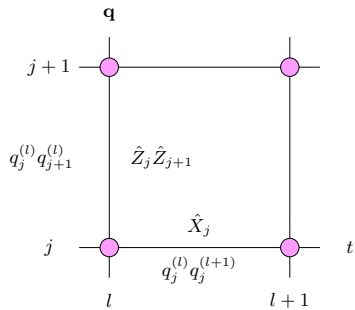
This physical interpretation is consistent with Lagrangian formalism when the Suzuki path integral is applied to this 1D transverse Ising chain style Hamiltonian (cf. \cref{sec:suzuki} Euclidean version).
The action corresponding to \cref{eq:ring_disagree} is the classical Hamiltonian of (1+1)D Ising model
(see \cref{fig:plaquette})
\begin{align*}
\action [\vec{\bm{\sigma}}]
	&= \sum_l \sum_j -h' \sigma^{(l)}_{j}\sigma^{(l+1)}_{j}-J'\sigma^{(l)}_j\sigma^{(l)}_{j+1}
	\\
	&\simeq 
	\sum_l \sum_j \frac{h'}{2} \qty(\sigma^{(l)}_{j}-\sigma^{(l+1)}_{j})^2
	+
	\frac{J'}{2}\qty(\sigma^{(l)}_j-\sigma^{(l)}_{j+1})^2 
\end{align*}
where $\pz_j^{(l)}\ket{q_j}=\sigma_j^{(l)}\ket{q_j}$.
$J'$ and $h'$ are independent of configurations $\ket{\vbq}$ but dependent on the interpolation function $\lambda$.
The first term in the action takes the form of $\integer_2$ version classical kinectic energy $1/2 mv^2$, 
while the second term $\integer_2$ version of harmonic (spring) potential $1/2 k (x_{j}-x_{j+1})^2$. 
From viewpoint of computer science,
$\sum_j\qty(\sigma^{(l)}_{j}-\sigma^{(l+1)}_{j})^2$ is the Hamming distance between two Trotter time steps (binary strings).
In consequence, the action has the physical interpretation of an ``arc length", which agrees with our intuition.
Meanwhile, we know that the runtime of an adiabatic algorithm is determined by the spectral gap of the quantum Hamiltonian. This fact can also be understood by the classical-quantum correspondence derived from path integral: a finite temporal correlation length corresponds to a spectral gap of the quantum problem \cite{sondhiContinuousQuantumPhase1997}.
(implied by the principle of uncertainty)
Unfortunately, we are not able to say how generically this intution holds.

\subsection{Optimize the protocol via Lagrangian}
The adiabatic theorem tells us that the evolution time $T$ of the QAA depends on the spectral gap $g(\hhat)$.
A commonly cited condition for adiabaticity \cite{rolandQuantumSearchLocal2002} is that
\begin{equation}
\max_t\frac{\norm{\dv{\hhat(t)}{t}}}{g^2(\hhat(t))}
	\le
	\dv{\lambda}{t}
	\frac{\max_\lambda\norm{\dv{\hhat(\lambda)}{\lambda}}}{\min_\lambda g^2(\hhat(\lambda))}
	\le \epsilon
\label{eq:adiabatic_condition}
\end{equation}
where the error tolerance $\epsilon\ll 1$ quantifies how the evolved state is close to the desired ground state.
However, the spectral gap is a priori unknown, 
obtaining knowledge about the spectral gap of a generic Hamiltonian can be as hard a problem as finding the ground state itself.

Grover's search is an excellent testbed for discussion on complexity of the heuristics
because it has known spectrum and the well-known quadratic quantum-classical complexity separation.
In the QAA version of Grover's search, the ground state of the final (problem) Hamiltonian $\hat{V}_G$ is the marked (but unknown) state $\ket{w}$,
while the initial (mixer) Hamiltonian $\hat{K}_G$ is diagonal in the Hadamard basis and has ground state $\ket{s}$ \cite{farhiAnalogAnalogueDigital1998}, namely,
\begin{equation}
\hat{K}_G
= \identity - \op{s}{s} 
,\quad
	\hat{V}_G
	= \identity - \op{w}{w}.
\label{eq:qaa_grover_hamiltonian}
\end{equation}
It is easy to check that the unitaries defined in \cref{eq:grover_diffusion_unitary} and \cref{eq:grover_oracle_unitary} are  generated by $\hat{K}_G$ and $\hat{V}_G$ respectively, as $\U_D=e^{-\ii\pi\hat{K}_G}$ and $\oracle_w = e^{-\ii\pi\hat{V}_G}$.
Analogous to \cref{eq:ring_disagree} which are kinetic and potential operators defined on hypercubes, 
$\hat{V}_G$ is the (discrete) Dirac delta potential to be minimized, while $\hat{K}_G$ is the adjacency matrix (discrete Laplacian) of a complete graph.

\begin{figure}[!ht]
	\centering
	\includegraphics[width=.85\linewidth]{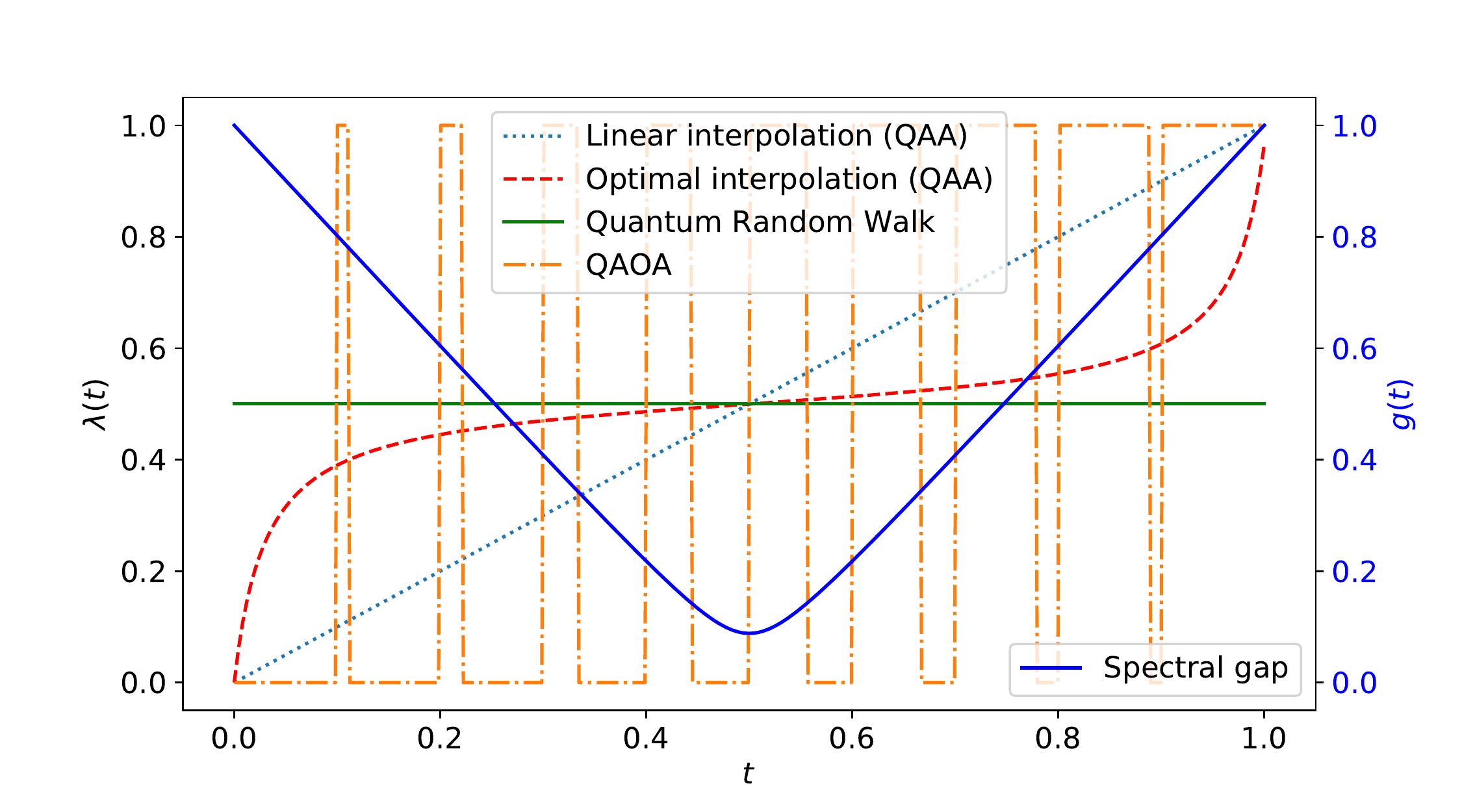}
	\caption{The protocols of different formulations of Grover’s search with $N=128$. The spectral gap is plotted with the blue solid line.}
\label{fig:qaoa}
\end{figure}

Given the Hamiltonians \cref{eq:qaa_grover_hamiltonian}, we can find the spectral gap of the interpolated Hamiltonian \cref{eq:adiabatic} 
\begin{equation}
	g(\lambda) = 
	\sqrt{1- 4\frac{N-1}{N} \lambda(1-\lambda)}
\end{equation}
in terms of $\lambda$
and $\norm{\dv{\hhat}{\lambda}}\le 1$.
Clearly, the minimum gap is $1/\sqrt{N}$ at $\lambda=1/2$.
If we try the linear interpolation $\lambda(t):=t/T$,
\cref{eq:adiabatic_condition} becomes 
\begin{equation}
	T 
\ge \max\frac{\norm{\dv{\hhat}{\lambda}}}{\epsilon g^2(\lambda)}.
\end{equation}
Thus, the evolution time $T$ is proportional to $N$.
That is to say, the linear interpolation doesn't exhibit speedup over classical computation.
Unsurprisingly, the interpolation $\lambda(t)$ can be improved to reach the optimal evolution time $\bigO(\sqrt{N})$ by applying adiabatic condition \cref{eq:adiabatic_condition} locally instead of globally, i.e., choosing the (\emph{ansatz}) interpolation $\dv{\lambda}{t}:=\epsilon g^2$ 
\cite{rolandQuantumSearchLocal2002} 
\cite{vandamHowPowerfulAdiabatic2001}. 
Specifically, we slow down the variation of $\lambda$ when the gap is small (center part in \cref{fig:qaoa}) and accelerate when the gap is large.
The total evolution time can be easily obtained by integration as
\begin{equation}
	T 
	= \int_0^1 \dv{t}{\lambda} \dd{\lambda} 
	= \frac{1}{\epsilon} \int_0^1 \frac{1}{g^2} \dd{\lambda} 
	= \frac{\pi}{2\epsilon} \sqrt{N}.
\label{eq:optimal_qaa}
\end{equation}

The above analysis can be recast in the language of Lagrangian formalism.
We reparameterize the interpolation function $\lambda$ by a set of control parameters $\vbx(t)=(x^{(1)}(t),\dots,x^{(M)}(t))$, (e.g., electric or magnetic fields, laser beams).
Then, the adiabatic time-functional is defined as integral of Lagrangian over $\lambda$
\begin{equation}
	\action[\vbx(\lambda)] = \int_0^1 \dd{\lambda} \dlagrangian(\vbx(\lambda),\dot{\vbx}(\lambda))
\end{equation}
where the Lagrangian $\dlagrangian(\dot{\vbx}(\lambda),\vbx(\lambda))=\norm{\dot{x}^i\partial_{x^i}\hhat(\vbx(\lambda))}/\epsilon g^2(\vbx(x))$ \cite{rezakhaniQuantumAdiabaticBrachistochrone2009}.
The optimal path $\vbx(\lambda)$ (Quantum Adiabatic Brachistochrone) should obey the Euler-Lagrange equation.
This argument is similar to Nielsen’s claim \cite{nielsenGeometricApproachQuantum2005} about the optimal synthesis of quantum circuits: determining an optimal quantum circuit is equivalent to finding the shortest path between two points in a certain curved geometry (geodesics equation).

The optimal parameters of QAOA version Grover's seach is found by \cite{jiangNearoptimalQuantumCircuit2017} based on spin coherent state formulation.
A handful of recent papers \cite{schifferAdiabaticSpectroscopyVariational2021} pay attention to the hybrid protocols like bang-anneal-bang protocols,
that is, combining annealing schemes (QAA) and variational schemes (QAOA) to get globally optimal performance.
The optimization of protocol costs is studied by classical methods such as Lagrange multiplier and Pontryagin's minimum principle (optimal control) \cite{yangOptimizingVariationalQuantum2017}.
It is worth noting that cost of the protocol is written as action in \cite{bradyOptimalProtocolsQuantum2021}.  

\section{Path Integral and Quantum Random Walks}\label{sec:walk}
Besides the circuit model and QAA discussed above, quantum random walk, the direct analogue of classical random walk,
is another universal paradigm of quantum computation 
that has shown advantages over its classical counterparts.
Quantum random walks have close relations with path integral and Lagrangian.
It is natural to formulate the continuous-time quantum random walk \cite{childsExampleDifferenceQuantum2002} with path integral
since it just generalizes the free quantum propagator to graphs (discrete space).
On the other hand, the distribution of the discrete-time quantum random walk \cite{nayakQuantumWalkLine2000} can be calculated by counting signed paths.
Moreover, the discrete-time quantum walk on a line takes the form of Dirac equation in continuum limit.

Before discussing the quantum random walk, it is worth giving a short historic review on classical random walk and its variants.
Random walks are a fundamental ingredient of nondeterministic algorithms which have wide applications.
Firstly, consider the simplest random walk, that is,
the symmetric discrete-time random walk on a one-dimensional lattice  $\integer$ (discrete-space).
The recurrence relation (master equation) of the probability distribution $\phi_j^{(l)}$ at time step $l$ and $j$ site is simply
\begin{equation}
    \phi_j^{(l+1)} = \frac{1}{2} \qty(\phi_{j+1}^{(l)} + \phi_{j-1}^{(l)} )
\label{eq:random_walk_recurrence}
\end{equation}
\begin{figure}[!ht]
    \centering
    \begin{subfigure}{0.49\textwidth}
    \centering
		\scalebox{0.7}{\begin{tikzpicture}
	\begin{pgfonlayer}{nodelayer}
		\node [style=none] (13) at (0, 0) {};
		\node [style=none] (14) at (-4, 0) {};
		\node [style=none] (15) at (-8, 0) {};
		\node [style=none] (17) at (-6, 0) {};
		\node [style=none] (18) at (-2, 0) {};
		\node [style=none] (19) at (0, 2) {};
		\node [style=none] (20) at (-4, 2) {};
		\node [style=none] (21) at (-8, 2) {};
		\node [style=none] (23) at (-6, 2) {};
		\node [style=none] (24) at (-2, 2) {};
		\node [style=none] (25) at (0, 0) {};
		\node [style=none] (26) at (-4, 0) {};
		\node [style=none] (27) at (-8, 0) {};
		\node [style=none] (29) at (-6, 0) {};
		\node [style=none] (30) at (-2, 0) {};
		\node [style=none] (31) at (0, 6) {};
		\node [style=none] (32) at (-4, 6) {};
		\node [style=none] (33) at (-8, 6) {};
		\node [style=none] (35) at (-6, 6) {};
		\node [style=none] (36) at (-2, 6) {};
		\node [style=none] (37) at (0, 8) {};
		\node [style=none] (38) at (-4, 8) {};
		\node [style=none] (39) at (-8, 8) {};
		\node [style=none] (41) at (-6, 8) {};
		\node [style=none] (42) at (-2, 8) {};
		\node [style=none] (66) at (6, 0) {};
		\node [style=none] (67) at (2, 0) {};
		\node [style=none] (68) at (4, 0) {};
		\node [style=none] (69) at (8, 0) {};
		\node [style=none] (71) at (6, 2) {};
		\node [style=none] (72) at (2, 2) {};
		\node [style=none] (73) at (4, 2) {};
		\node [style=none] (74) at (8, 2) {};
		\node [style=none] (76) at (6, 0) {};
		\node [style=none] (77) at (2, 0) {};
		\node [style=none] (78) at (4, 0) {};
		\node [style=none] (79) at (8, 0) {};
		\node [style=none] (81) at (6, 6) {};
		\node [style=none] (82) at (2, 6) {};
		\node [style=none] (83) at (4, 6) {};
		\node [style=none] (84) at (8, 6) {};
		\node [style=none] (86) at (6, 8) {};
		\node [style=none] (87) at (2, 8) {};
		\node [style=none] (88) at (4, 8) {};
		\node [style=none] (89) at (8, 8) {};
		\node [style=none] (90) at (-9.5, 0) {$4$};
		\node [style=none] (91) at (-9.5, 2) {$3$};
		\node [style=none] (92) at (-9.5, 4) {$2$};
		\node [style=none] (93) at (-9.5, 6) {$1$};
		\node [style=none] (94) at (-9.5, 8) {$0$};
		\node [style=none] (95) at (-8.75, 8) {};
		\node [style=none] (96) at (-8.75, 6) {};
		\node [style=none] (97) at (-8.75, 4) {};
		\node [style=none] (98) at (-8.75, 2) {};
		\node [style=none] (99) at (-8.75, 0) {};
		\node [style=none] (101) at (9, 8) {};
		\node [style=none] (102) at (9, 6) {};
		\node [style=none] (103) at (9, 4) {};
		\node [style=none] (104) at (9, 2) {};
		\node [style=none] (105) at (9, 0) {};
		\node [style=bold vertex] (109) at (0, 8) {};
		\node [style=bold vertex] (110) at (-2, 6) {};
		\node [style=bold vertex] (111) at (2, 6) {};
		\node [style=bold vertex] (112) at (0, 4) {};
		\node [style=bold vertex] (113) at (-4, 4) {};
		\node [style=bold vertex] (114) at (4, 4) {};
		\node [style=bold vertex] (115) at (6, 2) {};
		\node [style=bold vertex] (116) at (2, 2) {};
		\node [style=bold vertex] (117) at (-2, 2) {};
		\node [style=bold vertex] (118) at (-6, 2) {};
		\node [style=bold vertex] (119) at (-8, 0) {};
		\node [style=bold vertex] (120) at (-4, 0) {};
		\node [style=bold vertex] (121) at (0, 0) {};
		\node [style=bold vertex] (122) at (4, 0) {};
		\node [style=bold vertex] (123) at (8, 0) {};
		\node [style=none] (125) at (-4, -1) {$4/16$};
		\node [style=none] (126) at (4, -1) {4/16};
		\node [style=none] (127) at (-9.5, 9) {$t$};
		\node [style=none] (128) at (0, -1) {$6/16$};
		\node [style=none] (129) at (8, -1) {$1/16$};
		\node [style=none] (130) at (-8, -1) {$1/16$};
		\node [style=none] (133) at (10, 8) {$x$};
		\node [style=none] (134) at (0, 8.75) {};
		\node [style=none] (135) at (0, 8.75) {$0$};
		\node [style=none] (136) at (0, 8.75) {};
		\node [style=none] (137) at (2, 8.75) {$1$};
		\node [style=none] (138) at (4, 8.75) {$2$};
		\node [style=none] (139) at (-2, 8.75) {$-1$};
		\node [style=none] (140) at (-4, 8.75) {$-2$};
		\node [style=none] (141) at (6, 8.75) {$3$};
		\node [style=none] (142) at (8, 8.75) {$4$};
		\node [style=none] (143) at (-6, 8.75) {$-3$};
		\node [style=none] (144) at (-8, 8.75) {$-4$};
		\node [style=none] (145) at (10.25, 0) {$\phi(t,x)$};
		\node [style=none] (146) at (-4, -2) {$10/16$};
		\node [style=none] (147) at (4, -2) {2/16};
		\node [style=none] (148) at (0, -2) {$2/16$};
		\node [style=none] (149) at (8, -2) {$1/16$};
		\node [style=none] (150) at (-8, -2) {$1/16$};
		\node [style=none] (151) at (-10.5, -1) {Classical};
		\node [style=none] (152) at (-10.5, -2) {Quantum};
	\end{pgfonlayer}
	\begin{pgfonlayer}{edgelayer}
		\draw (39.center) to (27.center);
		\draw (41.center) to (29.center);
		\draw (38.center) to (26.center);
		\draw (42.center) to (30.center);
		\draw (37.center) to (25.center);
		\draw (87.center) to (77.center);
		\draw (88.center) to (78.center);
		\draw (86.center) to (76.center);
		\draw (89.center) to (79.center);
		\draw (95.center) to (101.center);
		\draw (96.center) to (102.center);
		\draw (97.center) to (103.center);
		\draw (104.center) to (98.center);
		\draw (105.center) to (99.center);
		\draw [style=bold edge] (109) to (111);
		\draw [style=bold edge] (111) to (112);
		\draw [style=blue edge] (109) to (110);
		\draw [style=blue edge] (110) to (112);
		\draw [style=blue edge] (112) to (117);
		\draw [style=red edge] (117) to (120);
		\draw [style=bold edge] (110) to (113);
		\draw [style=bold edge] (113) to (117);
		\draw [style=bold edge] (113) to (118);
		\draw [style=bold edge] (118) to (120);
	\end{pgfonlayer}
\end{tikzpicture}}
		\caption{Random walk on 1D lattice: classical one results in a Gaussian distribution; Discrete-time quantum walk with the initial state $\ket{0}\ket{\downarrow}$.}
\end{subfigure}
    \begin{subfigure}{0.49\textwidth}
    \centering
\includegraphics[width=.95\linewidth]{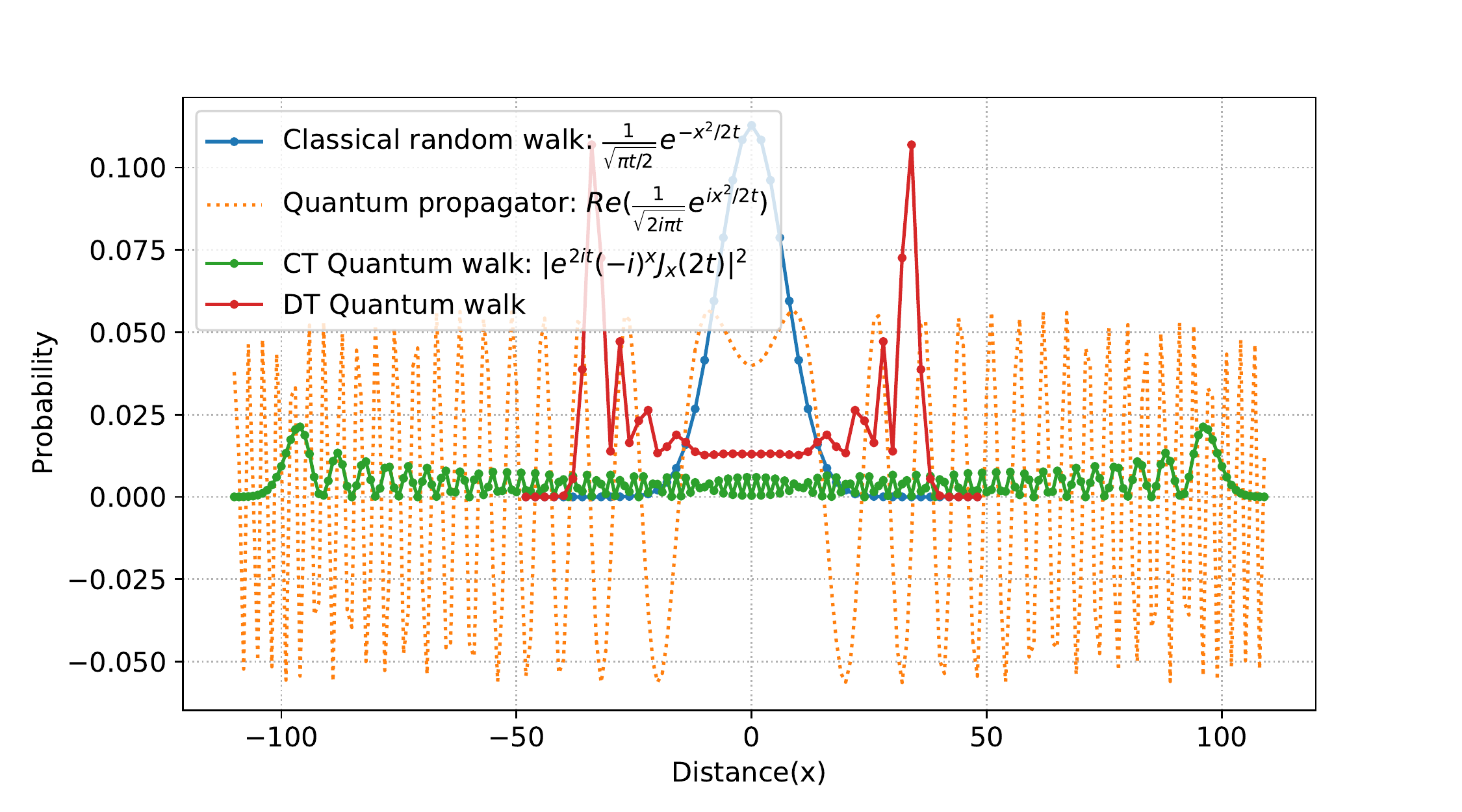}
		\caption{The number of coin tosses (time steps) is 50. DTQRW is symmetrized by the initial coin state $(\ket{0}+\ii\ket{1})/\sqrt{2}$. Only the real part of quantum propagator is plotted because its modulus is constant with fixed time.}
\end{subfigure}
    \caption{Different kinds of walks in 1d space.}
	\label{fig:walk}
\end{figure}
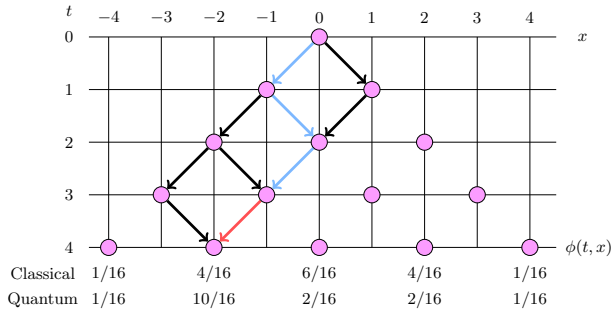
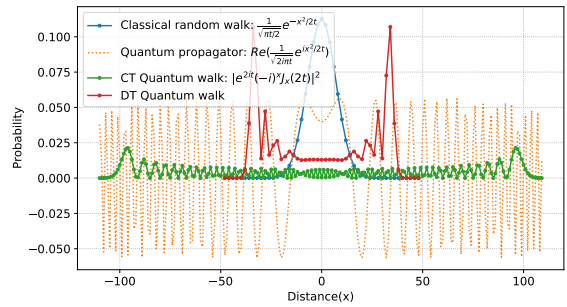
For this case, the probability distribution $\phi_j^{(n)}=\tbinom{n}{(n+j)/2}\cdot 2^{-n}$  can be easily sovled by combinatorial arguments:
the particle at time $n$ and $j$ site need to have $(n+j)/2$ heads from $n$ coin toss.
By the central limit theorem, as $l\to \infty$, 
the discrete probability distribution approaches a normal (Gaussian) distribution
\begin{equation}
	\phi(t,z) = 
\frac{1}{\sqrt{2\pi t/4}} e^{-z^2/(2t)}
	\label{eq:normal}
\end{equation}
which is the fundamental solution (heat kernel/Green function) of the heat (diffusion) equation (continuous version of \cref{eq:random_walk_recurrence} in both space and time)
\begin{equation}
    \pdv{\phi (t,z)}{t}
    =\pdv{ ^2 \phi (t,z)}{z^2}
\label{eq:heat}
\end{equation}
where $\phi$ is a scalar (continuous) field. 
Interestingly, the first path integral in history, i.e., Wiener process (Brownian motion) closely related with heat equation, inspired Feynman's path integral for quantum mechanics.

\subsection{Quantum propagator on graphs and continuous-time quantum walk}
In path integral formalism of quantum mechanics, all dynamics information is encoded in quantum propagator also called kernel.
Childs, Farhi, and Gutmann \cite{childsExampleDifferenceQuantum2002}
generalized classical (continuous time and space) diffusion equation (Eq. \ref{eq:heat}) to continuous-time quantum random walk (CTQRW) on discrete (space) graph.
It is also the quantum free propagation on graph (discrete space).
For the free propagator, we take the potential $V(x)=0$.
By direct analogy, the free Hamiltonain in \schrodinger equation \cref{eq:evolution} plays the role of discrete Laplacian in diffusion equation \cref{eq:heat}.
\begin{equation}
	\llaplacian:=\hat{A}-\hat{D}	
	\Longleftrightarrow
	\hhat_0=\frac{\phat^2}{2m}
=-\frac{\hbar^2\nabla^2}{2m}
\end{equation}
where $\hat{A}$ is the adjacency matrix of the graph and $\hat{D}$ is its diagonal degree matrix.
For the regular graphs (e.g. ring, hypercube, complete graph), the constant degree matrix $\hat{D}$ can be dropped.

The simplest case is the free propagation on a one-dimensional lattice,
where the Laplacian (adjacency) matrix is just the second order finite difference matrix.
Since there is no non-commutative term in the Hamiltonian, 
we don't need the time slice step \cref{eq:time_slice}.
Instead, we can directly compute this (discrete) free quantum propagator $\mel{z_F}{e^{-\ii t \llaplacian}}{z_I}$ by inserting a complete set of Fourier basis $\sum_{p}^N \op{\tilde{p}}{\tilde{p}}$.
Firstly, let's examine the eigenvalue of $\llaplacian$ in terms of its eigenstate 
$\ket{\tilde{p}}=\sum_{p}^N e^{\ii 2\pi pz/N} \ket{z}$ 
(the Fourier transform of $\ket{z}$ cf. \cref{eq:qft_zN}),
\begin{equation}
    \mel{z'}{\llaplacian}{\tilde{p}}
= 2\qty(\cos(\frac{2\pi}{N}p) -1) \ip{z'}{\tilde{p}}
\end{equation}
We would like to stress that $\cos(p) -1$ cannot be approximated as $p^2$ even for large $N$ because $p\in [N]$.
If we assume $N$ is large, the propagator can be approximately evaluated by replacing the summation with an integral
\cite[Sec 3.3.2]{childsQuantumInformationProcessing2004}
\begin{align}
	\mel{z_F}{e^{-\ii t\hhat_0 }}{z_I}
	&=\sum_{p=1}^{N} 
	e^{-\ii t 2\cos(\frac{2\pi}{N}p) +\ii \frac{2\pi}{N} p(z_I-z_F)} 
	\\
	&
\approx e^{2\ii t} (-\ii)^{d} J_{d} (2t)
	\label{eq:ctqrw_kernel}
\end{align}
where $J_d(2t)$ is the Bessel function and $d:=z_F-z_I$ is the distance between initial and final position.
In \cref{fig:walk}(b), we can see the significantly different behaviors between CTQRW on one-dimensional lattice and quantum free propagator in one-dimensional continuous space: the former one evolves as a wave packet with speed 2, the later one is a complex number with the modulus determined by time $t$ (independent of distance).

Furthermore, the CTQRW can be generalized to the case 
where the Hamiltonian contains a potential term instead of just a kinetic term (Laplacian).
The most celebrated example is Grover's search
\cite{farhiAnalogAnalogueDigital1998} 
\cite{childsSpatialSearchQuantum2004}
\begin{equation}
    \hhat 
= - \gamma \llaplacian - \hat{V}_G
	= - \gamma N \op{s}{s} - \op{w}{w} 
    \label{eq:grover_walk_hamiltonian}
\end{equation}
where $\gamma$ is a parameter to be tuned.
When $\gamma=1/N$ and $N$ is large, the quantum walk reaches the optimal running time $T=\pi \sqrt{N}/2$.
Compared with \cref{eq:qaa_grover_hamiltonian} and \cref{eq:adiabatic}, \cref{eq:grover_walk_hamiltonian} is just the case with $\lambda(t)=1/2$.
However, different from QAA always staying on the ground state, quantum walk version of Grover's search can overshot.

\subsubsection{Traveling salesman problem: quantum walk on weighted graphs}
In the preceding discussion, we restrict our attention on unweighted graphs.
Actually, it is very natural to consider quantum walk on weighted graphs.
The most well-known \nameref{def:np}-hard problem defined on weighted graph is the traveling salesman problem (\nameref{def:tsp}), 
which is to find the least weight Hamiltonian cycle of a weighted graph.
A weighted graph can be fully specified by a weight matrix (generalized Laplacian) $\hat{W}$ where the entries $\hat{W}_{ij}$ is the weight of the edge $(v_i,v_j)$.
The action (objective) is naturally formulated as the sum of weights of edges (Lagrangian) along one Hamiltonian cycle
\begin{equation}
	\zpartition \overset{?}{=}\sum_{z} \mel{z}{e^{-\beta\hat{W}}}{z}
	,\;
	\eaction 
= \sum_{l}^N \mel{z^{(l+1)}}{\hat{W}}{z^{(l)}}
\end{equation}
where $N$ is the number of vertices.
The cycle condition is automatically satisfied with periodic boundary condition required by the definition of partition function and we can insert $N$ complete basis to make sure $N$ vertices visited.
The tricky part is how to guarantee every vertex is visited exactly once.
There are several potential solutions:
the first one is setting the path constraint $\delta(permuate[N])$ by classical gates cf. \cref{eq:toffoli};
or by algorithmic (algebraic) ways such as linear programming, Lagrange multiplier;
or physical method: self-avoidance walk.

The above algorithm is just a proposal, might not be practically useful.
However, the principle of least action as a physical (natural) Lagrange multiplier solver should not be ignored.
It remains an interesting problem to propose a related quantum algorithm for (constrained) combinatorial optimization problems:
encoding the objectives and constraints into Lagrangian/action.
Evolution of the system generates a coherent ``weighted" sum over all possible configurations (paths).
Special properties of graphs could induce interference such that the paths around the stationary point (least action) canceled (or by semiclassical approximation).
Different from action in \autoref{sec:qaoa} quantifing the running time of quantum optimization algorithms, action defined here is an objective to optimize.
This kind of problem is not manifestly useful, 
but it might be useful to demonstrate quantum supremacy.

Quantum Hamiltonian complexity \cite{gharibianQuantumHamiltonianComplexity2015} is an extremely interesting topic of quantum computation which directly connects quantum physics and computer science.
Within our Lagrangian formalism of quantum computation, we expect its Lagrangian counterpart (Lagrangian/action complexity).
As we know, Kitaev, inspired by ideas due to Feynman, defined (finding ground state of) local Hamiltonian problem as the quantum analogue of \nameref{def:np}-complete problem \nameref{def:sat} and showed it is \nameref{def:qma}-hard \cite{kempeComplexityLocalHamiltonian2005}.
It is natural to ask what the typical QMA problem in path integral formalism is.
The quantum analogue of \nameref{def:tsp} is a suitable candidate, but a formal definition is unclear.

\subsection{Discrete-time quantum walk and Dirac Lagrangian}\label{sec:discrete_time}
As opposed to continuous-time quantum walk,
Andris Ambainis et al. \cite{ambainisOnedimensionalQuantumWalks2001} proposed the quantum analgue of discrete-time simple random walk (DTQRW).
Different from CTQRW only need position space, unitarity requires DTQRW to have extra coin space except for position space, $\hilbertspace_{position}\otimes \hilbertspace_{coin}$.
For the purposes of discussion, it is enough to limit ourselves to the one-dimensional line (lattice).
The quantum coin is realized by Hadamard gate $\U_{\hdm}$ acting on the coin space $\qty{\ket{\uparrow}\equiv\ket{0}\equiv\ket{\textup{R}},\ket{\downarrow}\equiv\ket{1}\equiv\ket{\textup{L}}}$.
After tossing the quantum coin, the coin state is in a superposition of ``turn left" and ``turn right".
Accordingly, one step of DTQRW is finished by a conditional/controlled translation operation $\U_{cT}$:
\begin{equation}
    \U_{cT}:
	\ket{z} \ket{q} \mapsto \ket{z+e^{\ii \pi q}} \ket{q}
	,\; q\in \qty{0,1},
\end{equation}
moving a particle one step left or right according to the coin state $\ket{q}$.
Technically, we carry this out by $\U_{cT}=e^{-\ii \phat \otimes \hhat} = \sum_k e^{-\ii E_k \phat} \otimes \op{E_k}{E_k}$.
This is discrete Von Neumann measurement \cite{childsQuantumInformationProcessing2004} with $\hhat=\pz$ with $E_k=\pm 1$,
also an instance of quantum phase estimation (an important application of quantum Fourier transform).
As shown in \cref{fig:walk}(b), the discrete-time quantum walk has the peak around $t$, while the peak of continuous-time counterpart is at $2t$ (quantum tunneling).

Another way to incorporate the degree freedom of coin is using
two-component spinor (chiral) representation 
$\psi(n,z):=\qty(\psi_{\textup{L}},\phi_{\textup{R}})^{\T}$.
The amplitudes at position $z$ after $n$ steps (toss coin) can be analytically obtained by combinatorial method.
Observe the colored path in \cref{fig:walk}(a), we should notice the fact that two consecutive left moves would induce a $(-1)$ phase to the amplitude.
Let the number of left moves among $n$ moves denoted by $n_l:= \frac{n-z}{2}$ and right moves $n_r:= \frac{n+z}{2}$,
then the left component of amplitude is expressed as 
\begin{equation}
	\psi_{\textup{L}}(n,z) = 
	\frac{1}{\sqrt{2^n}} \sum_b
	\binom{n_l-1}{b}
	\binom{n_r}{b}
	(-1)^{n_l-b-1}
	\label{eq:combinatorial_walk}
\end{equation}
with the right component computed simiarly.
The combinatorial explanation of \cref{eq:combinatorial_walk} is that firstly choosing $b$ intervals from $n_l-1$ intervals and then distribut $n_r$ right moves into these $b$ intervals.
The phase of a path is determined by the number of consecutive left moves.
Thus, the sums of binomial coefficients in \cref{eq:combinatorial_walk} are discrete counterparts of path integrals.

Some work \cite{meyerQuantumCellularAutomata1996} 
\cite{chandrashekarRelationshipQuantumWalk2010}
have shown the relativistic character of DTQRW.
Actaully, this idea goes back to the discrete version of the one-dimensional Dirac equation propagator (Feynman's checkerboard) considered by Feynman and Hibbs \cite[problem 2-6]{feynmanQuantumMechanicsPath2010}:
a particle of mass $m$ moving in one-dimensional lattice of spacing $a$ can go only forward or backward one step at the velocity of light $c$.
The kernel of the particle arriving at position $z$ is
(we keep the Feynman's original notations)
\begin{equation}
	K(z) = \sum_R N(R) (\ii a m)^R 
	\label{eq:feynman_chessboard}
\end{equation}
where $R$ is the number of reverses (corners) along a path and $N(R)$ denotes the number of paths that has $R$ reverses. 
Despite the slight difference bewtween the definitions of DTQRW and Feynman's checkerboard,
\cref{eq:feynman_chessboard} and \cref{eq:combinatorial_walk} have the same combinatorial nature: category all paths by the phase and count their times.
With Feynman's checkerboard description, it is natural to represent a $n$-step walk by a $n$-spin classical Ising chain in time direction cf. \cref{eq:classical_ising}.
With the key observation that $2R=\sum_l 1 - \sigma^{(l)}\sigma^{(l+1)}$ and $z=\sum_l^n\sigma^{(l)}$,
the correspondence between Feynman's checkerboard and (1+1)-dimensional Dirac equation 
\begin{equation}
    \ii\hbar\pdv{t}\psi= \qty( -\ii\hbar \sz \pdv{z} + mc \sx ) \psi
\end{equation}
can be established by transfer matrix method \cite{gerschFeynmanRelativisticChessboard1981}.
The Dirac equation can be derived from the Lagrangian density
\begin{equation}
\dlagrangian_{\textup{Dirac}} = \ii\hbar c \bar{\psi} \gamma^{\mu}\partial_{\mu}\psi - mc^2 \bar{\psi}\psi
\end{equation}
where $\bar{\psi}:=\psi^{\dagger}\gamma^0$. 
For (1+1)D case, the gamma matrices are $\gamma^1=-\ii\sy$ and $\gamma^0=\sx$ such that $ \gamma^0\gamma^1 = \sigma_z$.
Essentially, the correspondence between DTQRW and Dirac Lagrangian suggests a method of quantum simulation based on path integral and Lagrangian.

\section{Quantum Simulation with Action}\label{sec:simulation}

\subsection{Path integral representations of quantum simulation and sign problems}
First of all, we need to clarify the terminology ``quantum simulation".
In the context of quantum computation, \emph{quantum simulation} is simulating physical systems by quantum (computer) algorithms instead of simulating quantum systems by classical algorithms.
Note that the systems simulated by quantum computers are not necessarily quantum but also can be classical systems described by computationally hard differential equations or combinatorial problems.
Such terminologies as Quantum Monte Carlo (QMC) methods are ambiguous to some readers because QMC methods are classical (computer) algorithms for simulating quantum (many-body) systems \cite{landauGuideMonteCarlo2015}, not quantum algorithms at least originally.
Although some classical algorithms for simulation have their quantum versions, 
we still need to be aware of the difference.

In current research, quantum simulation mainly refers to quantum Hamiltonian \emph{(real-time) dynamics} simulation. 
Precisely, given a quantum Hamiltonian $\hhat$, an initial state $\ket{\psi}$, and evolution time $t$, the goal is to produce the final state $e^{-\ii t\hhat}\ket{\psi}$ within error tolerance $\epsilon$.
There are two points worth noting:
(1) The default description of quantum evolution is \schrodinger equation, but there are at least two alternative descriptions: Heisenberg's picture and Feynman's path integral.
(2) Besides dynamics (real-time) simulation, there are \emph{finite temperature} (imaginary-time/static/equilibrium) simulation
of which the typical tasks are computing the static properties (e.g. correlation) and preparing Gibbs mixed state or ground state (zero-temperature).
So, in more general scenario, a generic quantum simulation problem can be defined to evaluate
\begin{equation}
	\Tr(e^{\ii t\hhat } \hat{Q}_1 e^{-\ii t\hhat}\hat{Q}_2 \hat{\rho})
	\label{eq:simulation}
\end{equation}
where $\hat{\rho}:= e^{-\beta \hhat}$ is the density matrix
and $\hat{Q}$ is certain observable of interest.
$e^{\ii t\hhat } \hat{Q}_1 e^{-\ii t\hhat}$ is the Heisenberg representation of real-time dynamics and the trace $\Tr$ means that it is statistical.
The good news is that both real- and imaginary-time evolution can be cast into in Lagrangian (path integral) formalism in the same manner, up to a Wick rotation. (see \cref{sec:path_integral})

The key motivation for Feynman's proposal \cite{feynmanSimulatingPhysicsComputers1982} for quantum computer is the belief that classical computers are bad at simulating highly entangled many-particle quantum (field) systems while quantum computers might be capable of doing it efficiently.
One manifest reason is classical digital simulation at least requires exponential (memory) space for keeping track of all coefficients of a $n$-qubit quantum state
while quantum computer only need linear space (number of qubits).
The tricky question is how to design polynomial time (efficient) quantum simulation algorithms.
For many realistic (non-frustrated) physical systems, (classical) path integral QMC algorithm is efficient and successful, 
but some Fermionic systems such as Hubbard model suffer from \emph{sign problem} which makes simulation intractable (exponential time in terms of the system size).
Specifically, if the path integral QMC is applied to real-time (dynamic) simulation,
the integrand $e^{\ii\action/\hbar}$ oscillates rapidly in high dimensional space.
While clever stationary-phase forms of the QMC method have been developed, 
acceptable solutions are available only for specific systems and possible only for relatively short times \cite{ortizQuantumAlgorithmsFermionic2001}. 
This kind of sign problem is called dynamical sign problem which has been alleviated by universal (circuit) quantum algorithms. 
Lloyd's seminal work \cite{lloydUniversalQuantumSimulators1996} showed that 
a quantum computer can efficiently reproduce the dynamical evolution of quantum (local Hamiltonian) system.
The basic idea of the scheme is decomposing the evolution $e^{-\ii t\hhat}$ into local terms by time slicing (Trotterization) where each local term can be simulated efficiently and the error is bounded by product formula \cref{eq:product_formula}.
This scheme's speedup over classical simulation comes from the locality (sparsity) of the Hamiltonians and parallelizability nature of quantum computation.
Except for the Lie-Trotter-Suzuki decomposition, other compilations of $e^{-\ii t\hhat}$ such as truncated Taylor series \cite{berrySimulatingHamiltonianDynamics2015}, Linear Combination of Unitaries \cite{childsHamiltonianSimulationUsing2012}, Qubitization (quantum walk) \cite{lowHamiltonianSimulationQubitization2019}, quantum signal processing (quantum singular value transformation) \cite{gilyenQuantumSingularValue2019} etc,
have been proposed to achieve optimal dependence on different parameters.

Another intuition for efficient quantum dynamics simulation is that unitary operations are cheap on quantum computers.
However, this scheme does not apply to imaginary-time simulation because it is not unitary.
Although $e^{-\beta E}$ is always positive,
negative terms in the sum over all ``paths" emerge when Euclidean path integral is applied \cite{troyerComputationalComplexityFundamental2005}.
This oscillation between positivity and negativity leads to exponential number of samples for simulation accuracy.
(minimal example of this \emph{statistical sign problem} cf. \cref{eq:sign_problem})
Many years of effort have been made to solve the sign problem, including classical and quantum algorithms. 
For example, the quantum version metropolis algorithm has been proposed to efficiently prepare the ground state or Gibbs state \cite{temmeQuantumMetropolisSampling2011} by utilizing (inverse) phase estimation.
It permits sampling directly from the eigenstates of the Hamiltonian, and evades the sign problem.
Despite these efforts, finding the ground state of 2-local Hamiltonian is proved to be hard in general even for quantum computers, i.e., this problem is \nameref{def:qma}-complete (quantum analogue of \nameref{def:np}-complete).
Note that QAA and QAOA mentioned in \cref{sec:qaoa} are the heuristics for finding the ground state of a diagonal Hamiltonian and only efficient for specific systems.

\subsection{Quantum simulation with action and path integral}
As there are digital and analog classical computation,
there are mainly two variants of quantum simulation schemes:
(1) Digital quantum simulation (DQS): decompose the unitary evolution into a set of basic (universal) gates. 
It is Lloyd's scheme mentioned before and a typical application is quantum circuit Hamiltonian simulation of quantum $\phi^4$ field theory \cite{jordanQuantumAlgorithmsQuantum2012}.
This scheme is scalable, but it need error correction and lacks physical meaning.
(2) Analog quantum simulation (AQS): 
one (easy-to-control) quantum system reproduces certain property of the other one (of interest)
\cite{georgescuQuantumSimulation2014}.
An important advantage of AQS is that it could be useful even in the presence of errors, up to a certain tolerance level. 
The simulation time is usually linear to the system evolution time because of the direct mapping between two systems' Hamiltonians.
(3) Nonetheless, the boundary between DQS and AQS is not so clear.
Some quantum algorithms were designed as analog ones,
but they are universal and written in circuit language.
AQS was believed to be non-universal, but recently Cubitt et al. \cite{cubittUniversalQuantumHamiltonians2017} have proposed the universal analog Hamiltonian.
There is also a hybrid of digital and analog one e.g. \cite{janeSimulationQuantumDynamics2002}, 
where the time evolution is Trotterized but the different terms of the Hamiltonian are implemented using an analog simulation instead of quantum gates.

Nevertheless, no matter digital or analog, real- or imaginary-time, the starting point of current mainstream quantum simulation schemes is Hamiltonian.
However, Hamiltonian is not always the most natural language, especially in gauge field simulation e.g.
QED \cite{stetinaSimulatingEffectiveQED2020} and
QCD \cite{shawQuantumAlgorithmsSimulating2020},
where the system is more conveniently described by its Lagrangian and action.
The common way to circumvent this problem is transforming the Lagrangian into its effective Hamiltonian.
This modification however brings its own set of caveats:
introducing a lattice breaks Lorentz invariance badly
such that the parameters should be carefully tuned \cite{preskillSimulatingQuantumField2018}.
In this paper, we are going to switch the starting point of quantum simulation from (digital) Hamiltonian to (analog) Lagrangian and find mappings (duality/correspondence) between Lagrangians.
The motivation is mainly threefold:
(1) Lagrangian (path-integral) formalism provides a unified framework for real- and imaginary-time evolution.
(2) Lagrangian formalism is the preferrable language in condensed matter physics (conformal field theory) and high energy physics (quantum field theory).
(3) Lagrangian formalism may be a suitable framework for analog quantum computing,
where it could be straightforward to find the physical implementation of the algorithms and simulation time is linear.

\subsubsection{Path integral: Duality between U(1) gauge theory and XY model}

The main challenges of a quantum simulation of lattice gauge field theories are threefold \cite{zoharQuantumSimulationsLattice2016}:
(1) contain both fermions and bosons
(2) be Lorentz invariant i.e., to have a causal structure
(3) have local gauge invariance, which is the symmetry responsible for gauge-matter interactions.
Quantum electrodynamics (QED) 
is the minimal model satisfying these three requirements.
The Lagrangian density of QED has three components: matter (Fermion), light (photon), and light matter interaction
\begin{equation}
    \dlagrangian_{\textup{QED}} =
\dlagrangian_{\textup{Dirac}} - e\bar{\psi}\gamma^{\mu}A_{\mu}\psi 
	- \frac{1}{4} F_{\mu\nu}F^{\mu\nu}
    \label{eq:qed_lagrangian}
\end{equation}
where the field strength $F_{\mu\nu}=\partial_{\mu}A_{\nu}-\partial_{\nu}A_{\mu}$ and vector field $\vb{A}$.

In \cref{sec:walk}, we have shown the correspondence between DTQRW and Dirac Lagrangian.
Here, we will see the duality between U(1) gauge theory (photon) and XY model.
One-dimensional classical XY model is described by
planar spin $\vb{s}_j = \qty(\cos(\theta_j), \sin(\theta_j))^{\T}$ placed on each lattice site $j$
and the nearest-neighbors interaction
\begin{equation}
	H 
	= \sum_j \vb{s}_j \cdot \vb{s}_{j+1}
	= \sum_j \cos (\theta_{j} - \theta_{j+1}).
\label{eq:xy_model}
\end{equation}
By the correspondence between quantum mechanics and statistical mechanics, the Euclidean Lagrangian of 2-dimensional quantum XY model is a classical Hamiltonian of (2+1)-dimenional XY model \cite{wallinSuperconductorinsulatorTransitionTwodimensional1994} \cite{sondhiContinuousQuantumPhase1997},
similar to \cref{eq:d_to_d+1_mapping}.
\cite{kogutIntroductionLatticeGauge1979} \cite{savitDualityFieldTheory1980} \cite{wenQuantumFieldTheory2010} show that the path integral of (2+1)D XY-model 
\begin{equation}
	\zpartition_{\textup{XY}} = \int \D\theta 
e^{\ii\int \dd{t}\dd^2{x} \dlagrangian_{\textup{XY}}}
	,\;
	\dlagrangian_{\textup{XY}} = \frac{\chi}{2} \qty(\dot{\theta}^2-(\partial_x\theta)^2)
	\label{eq:u1_gauge}
\end{equation}
and U(1) gauge theory describe the same physical system
\begin{equation}
	\zpartition_{\textup{U(1)}} = \int \D a_{\mu} e^{\ii\int \dd{t} \dd^2{x} \dlagrangian_{\textup{U(1)}}}
	,\;
	\dlagrangian_{\textup{U(1)}} = \frac{1}{2g^2} (\vb{e}^2-b^2)
\end{equation}
where $e_i = \partial_0 a_i-\partial_i a_0$ is the ``electric'' field and $b=\partial_1 a_2 - \partial_2 a_1$ is the ``magnetic'' field.

As we know, the experiments of particle physics are complicated and expensive, while the condensed matter systems are relatively easy to manipulate in labs.
With the mapping between the Lagrangians of two systems, we can obtain the information of certain system of interest such as QCD by doing the experiment of another physical system.
In this paper, we only show some evidence for our proposal. 
It remains to be formally extended to full QED in (3+1)D, even non-Abelian QCD. The ultimate goal is achieving simulation of the standard model by finding a set of ``universal Lagrangian gadgets” from condensed matter physics.

\section{Outlook and Discussion}\label{sec:discussion}
This paper demonstrates that various aspects of quantum computation, 
including complexity analysis and quantum simulation, 
can be studied within this Lagrangian (path integral) framework. 
At the same time, these results raise many interesting questions in different directions
and many advantages of Lagrangian formalism remain to be exploited, e.g.,
\begin{itemize}[leftmargin=*]
	\item In \autoref{sec:qaoa}, we have established the informal argument and intuition that the action converted from Hamiltonian by path integral plays the role of complexity (cost) of a quantum algorithm.  
	This belief is supposed to be more rigorous and formal such that more tools in quantum physics such as phase transition can be applied to the complexity analysis of quantum algorithms.
\item In \autoref{sec:review_circuit}, we rewrite quantum circuits with the language of Lagrangian formalism. 
	However, it would be interesting to investigate the universality of quantum computing with Lagrangian as starting point,
	i.e., proving Lagrangian version of Kitaev's (universal) theorem. 
\item Topological computing \cite{kitaevFaulttolerantQuantumComputation2003} is a promising model of quantum computation credited to its robustness to quantum errors (decoherence). 
Topological terms appear in spin coherent state path integral 
    which is an interesting question which we do not pursue in this paper.
    The study of topological quantum computing through path integral might shed light on the design of quantum algorithms and physical implementations of topological quantum computer.
\item Symmetry and invariance are extensively studied in elementary particle physics via group theory and Lagrangian formalism,
    while group theory is a useful tool for analyzing quantum algorithms (quantum Fourier transform) for algebraic problems \cite{childsQuantumAlgorithmsAlgebraic2010}.
	It is natural to ask whether fundamental symmetries of physics bound the power of quantum computation.
\end{itemize}

\bibliographystyle{apsrev4-2}
\bibliography{ref}

\onecolumngrid
\appendix

\section{Path Integral Formalism}\label{sec:path_integral}
In case that some computer science readers may not be familiar with path integral formalism that was developed in physics,
we give an introductory review of it at the beginning of the appendix.
In this section, we will go back and forth between discrete and continuous variables, real- and imaginary-time evolutions, quantum and classical systems.
You can skip \cref{sec:path_integral} if you are comfortable with path integrals in the main text.

One central formula used in path integral is the Gaussian integral.
First of all, the basic Gaussian integral for a real scalar variable $x$ reads 
$\intinf e^{-\frac{1}{2} ax^2} \dd{x}= \sqrt{2\pi/a}$ with $a\in \realnumber^+$,
and more generally
\begin{equation}
	\intinf e^{-\frac{1}{2} ax^2+bx} \dd{x}
	=\sqrt{\frac{2\pi}{a}} e^{\frac{b^2}{2a}}.
	\label{eq:gaussian_integral}
\end{equation}
Straightforwardly, this formula can be generalized to the complex case 
\begin{equation}
	\intinf e^{ -\frac{1}{2} ax^2+\ii bx } \dd{x}
	=\sqrt{\frac{2\pi}{a}} e^{-\frac{b^2}{2a}}
	,\quad
	\intinf e^{\ii \qty(-\frac{1}{2} ax^2+bx)} \dd{x}=\sqrt{\frac{2\pi}{\ii a}} e^{\ii\frac{b^2}{2a}},
\end{equation}
where the first one yields a new Gaussian function and the second one gives a function on the complex plane.

\subsection{Feynman's path integral for single particle systems (continuous variable)}\label{sec:path_integral_particle}
For simplicity, we shall begin our review of path integral formalism with the most introductory system, i.e., a single particle moving in a one-dimensinoal position-dependent potential, described by a (time-independent) quantum Hamiltonian $\hhat=\phat^2/2m+V(\xhat)$.
The Hilbert space is spanned by infinite (continuous) coordinate states $\ket{x}$, the eigenstates of position operator $\xhat$.
The quantity that we are interested in is quantum propagator or called kernel that encodes all dynamic information, namely, $\mel{x_F}{e^{-\ii t\hhat /\hbar}}{x_I}$
where $x_I,x_F$ are the initial and final positions respectively.
\footnote{Quantum propagator has various equivalent notations i.e. $K(x_F,t_F;x_I,t_I) \equiv U(x_F,t_F;x_I,t_I)\equiv \ip{x_F,t_F}{x_I,t_I} \equiv (x_F,t_F|x_I,t_I) \equiv G(x_F,x_I;t):=\mel{x_F}{e^{-\ii t\hhat /\hbar}}{x_I}$.}

\subsubsection{Tricks: time slice, product formula and complete basis}
Brute-force calculation of quantum propagator is hard in general on classical computers because the matrix exponentiation of Hamiltonian $e^{-\ii t \hhat}$ entails manipulation of matrices with intractable sizes.
The first trick of Feynman's path integral is slicing the Hamiltonian into infinite small time intervals and inserting the complete set of position basis $\int\dd{x}\op{x}{x}=\identity$ at each time step
\begin{align}
    \mel{x_F}{e^{-\ii t\hhat }}{x_I}
    =& 
    \langle x_F |
    e^{-\ii \deltat \hhat }
    e^{-\ii \deltat \hhat }
    \cdots
    e^{-\ii \deltat \hhat }
    | x_I \rangle
=
    \qty(
        \prod_{l=1}^{M-1}
        \int \dd x^{(l)}
    )
\mel{x^{(l)}}{ e^{-\ii \deltat \hhat } }{x^{(l-1)}}
\label{eq:time_slice}
\end{align}
such that we can apply the second trick (product formula).
Since the position operator $\xhat$ and momentum operator $\phat$ don't commute,
the kinetic term and potential term don't commute.
Suzuki-Lie-Trotter product formula (or Campbell-Baker-Hausdorff theorem)
allows us to separate the kinetic and potential as
\begin{equation}
	e^{\hat{K}+\hat{V}} 
	= \qty(e^{(\hat{K}+\hat{V})/M})^M
	= \lim_{M\to \infty}\qty(e^{\hat{K}/M} e^{\hat{V}/M})^M
\text{ with } \;
	\norm{e^{\hat{K}+\hat{V}} - e^{\hat{K}}e^{\hat{V}}}_2 
	\in \bigO\qty(\norm{\hat{K}\hat{V}}_2)
	\label{eq:product_formula}
\end{equation}
The approximation becomes exact in the continuum limit $M\to\infty$.
Now, we can evaluate the kinetic and potential individually by inserting another complete basis $\int \dd{p}/2\pi \op{p}{q} =\identity$, 
where $\ket{p}$ is the eigenvector of the momentum operator $\phat$ (the continuous Fourier transform of $\ket{x}$: $\ip{x}{p}=e^{\ii px}$).
Focus on an individual factor $\langle x^{(l+1)}| e^{-\ii \deltat (\hat{K}+\hat{V})}| x^{(l)}\rangle$, we have
\begin{align}
\mel{x^{(l+1)}}{ e^{-\ii \deltat \phat^2/2m} e^{-\ii\deltat \hat{V}(\xhat)} }{x^{(l)}}
=& 
\int \frac{\dd p^{(l)}}{2\pi}
    \mel{x^{(l+1)}}{ e^{-\ii \deltat \phat^2/2m} }{p^{(l)}} 
    \mel{p^{(l)}}{e^{-\ii\deltat \hat{V}(\xhat)} }{x^{(l)}}
\\  
    =& 
    \int \frac{\dd p^{(l)}}{2\pi}
    e^{-\ii \deltat (p^{(l)})^2/2m}
    \ip{x^{(l+1)}}{p^{(l)}}
    e^{-\ii \deltat V(x^{(l)})}
    \ip{p^{(l)}}{x^{(l)}}
\\
    =& 
    e^{-\ii \deltat V(x^{(j)})}
    \int \frac{\dd p^{(l)}}{2\pi}
    e^{-\ii \deltat (p^{(l)})^2/2m}
    e^{\ii p^{(l)} (x^{(l+1)}-x^{(l)})}
\label{eq:insert_basis}
\end{align}
where we have applied, respectively: the momentum basis is a resolution of identity; $\ket{x}$ is the eigenstate of $\xhat$ and $\ket{p}$ is the eigenstate of $\phat$; and $\ip{x}{p}=e^{\ii px}$.
If we stop at this step and plug \cref{eq:insert_basis} into \cref{eq:time_slice}, we get the path integral in phase space $(x,p)$:
\begin{equation}
	\mel{x_F}{e^{-\ii t \hhat}}{x_I}
    =	
	\frac{1}{(2\pi)^M} \qty(\prod_{l=1}^{M} \int \dd{x^{(l)}} \dd{p^{(l)}})
	\exp(\sum_{l}^{M}\ii \deltat 
	\qty(p^{(l)} \cdot \frac{x^{(l+1)}-x^{(l)}}{\deltat}-\frac{(p^{(l)})^2}{2m} 
	- V\qty(x^{(l)}))
	)
	\label{eq:phase_space_integral}
\end{equation}
Note that the expression in the exponential equals $p \dot{x} - H(p,x;t)$ in continuum limite $\Delta t\to 0$, known as \emph{Legendre transform} of Lagrangian.
Actually, we can take a one more step: integrate out momentum by applying Gaussian integral \cref{eq:gaussian_integral} to \cref{eq:insert_basis}, i.e.,
\begin{equation}
	\int \frac{\dd{p}}{2\pi} 
	\exp(-\ii \frac{\deltat}{2m} p^2 + \ii p \qty(x^{(l+1)}-x^{(l)}))
	=
    \qty(\frac{-\ii m}{2\pi \deltat})^{1/2}
    \exp(\ii\deltat \frac{m}{2} \frac{\qty(x^{(l+1)}-x^{(l)})^2}{\deltat^2}).
\label{eq:gaussian_path_integral}
\end{equation}
By putting \cref{eq:gaussian_path_integral} back to \cref{eq:insert_basis} and \cref{eq:time_slice}, we express the quantum propagator as a high-dimensional integral
\begin{equation}
    \mel{x_F}{e^{-\ii t\hhat }}{x_I}
    =
    \qty(\frac{-\ii m}{2\pi \deltat})^{M/2}
    \qty(
        \prod_{l=1}^{M-1} \int \dd x^{(l)}
    )
\exp(\sum_{l=0}^{M-1} \ii\deltat \qty(\frac{m}{2} \frac{\qty(x^{(l+1)}-x^{(l)})^2}{\deltat^2}-V(x^{(l)}))).
	\label{eq:path_integral_slice}
\end{equation}
We can now take the continuum limit $\deltat \to 0$ ($M\to \infty$):
\begin{equation}
    \frac{\qty(x^{(l+1)}-x^{(l)})^2}{\deltat^2}
    \to \dot{x}^2
    ,\quad
    \deltat \sum_{l=0}^{M-1}  \to \int_{t_I}^{t_F} \dd{t}
    ,\quad
\qty(\frac{-\ii m}{2\pi \deltat})^{M/2}
    \qty(
        \prod_{l=1}^{M-1} \int \dd x^{(l)}
    )
    \to \int \D x(t) 
\end{equation}
such that we finally obtain the elegant formula of path integral in coordinate space
\begin{align}
    \mel{x_F}{e^{-\ii t \hhat}}{x_I}
    =
    \int \D x(t)\,
    \exp(\ii \int_{t_I}^{t_F}\dd{t}\qty(\frac{1}{2} m \dot{x}^2-V(x) ) )
:=
    \int \D x(t)\,
    e^{\ii \action[x(t)]}.
\label{eq:path_integral}
\end{align}
The argument also applies to $d$-dimensional systems with generalized coordinate $\vbq(t)$
\begin{equation}
    \action [\vbq] = 
    \int_{t_I}^{t_F}
    \dlagrangian(\vbq(t),\dot{\vbq}(t);t) \dd{t}
    ,\;
	\lagrangian(\vbq(t),\dot{\vbq}(t);t):=\frac{1}{2}m\dot{\vbq}^2(t)-V(\vbq(t))
    ,\;
    \vbq(t): \realnumber^+ \to \realnumber^d
	\label{eq:minkowski_action}
\end{equation}
The time integral of the Lagrangian is called the (Minkowski space/real-time) action. 
The action $\action[\vbq]$ is a functional which takes a function (path) as input,
while the Lagrangian $\dlagrangian$ is a function evaluated by the generalized position and velocity at every time.

\subsubsection{Principle of least action and semiclassical approximation}
To restore classical mechanics from action, we determine the form of the extremum path where variation of action is zero
\begin{equation}
    \delta \action 
    = \int \delta \dlagrangian(\vbq,\dot{\vbq};t) \dd{t}
= 0
	\Longrightarrow
    \pdv{\lagrangian}{q_j}-\dv{t}\pdv{\lagrangian}{\dot{q}_j}=0
    \label{eq:euler_lagrange}
\end{equation}
where the differential equation is the classical equation of motion, called \emph{Euler-Lagrange equation}. 

It is difficult to find the analytical solution of a generic path integral, 
but the simplest case, i.e., free propagator $V(x)=0$, can be rigorously solved by inserting momentum eigenstates and applying Gaussian integral
\begin{equation}
\mel{x_F}{e^{-\ii t \phat^2/(2m\hbar) }}{x_I}
    =
    \sqrt{\frac{m}{2\pi \ii \hbar t}}
    \exp(\frac{\ii}{\hbar}\frac{m (x_F-x_I)^2}{2t}).
    \label{eq:free_propagator}
\end{equation}
It is easy to see that the exponent is exactly the action contributed by the classical motion, i.e., $\action[x_{\textup{C}}(t)]=\frac{m}{2}\frac{(x_F-x_I)^2}{t-0}$.
That is to say, the paths except the stationary path are all canceled by symmetry.
This implies a trick for approximately evaluating more complicated path integrals, called semiclassical approximation: expand the action around the stationary point.

\subsubsection{Euclidean (imaginary-time) path integral, partition function, statistical mechanics}\label{sec:euclidean_path_integral}
In statistical mechanics, almost all thermodynamics quantities of interest can be computed from knowledge of partition function $\zpartition$.
According to the fundamental assumption of statistical mechanics, the probability that a classical equilibrium system with classical Hamiltonian function $H$ found in certain configuration $\vb{q}$ is $\probability(\vb{q})=e^{-\beta H(\vb{q})}/\zpartition$,
where $\beta:= 1/(k_B T)\in \realnumber^{+}$ is the inverse of temperature and the normalization factor $\zpartition:=\sum_{\qty{\vb{q}}} e^{-\beta H(\vb{q})}$ called partition function is the sum of Boltzman weights over all configurations. 
Certain logarithmic derivatives of $\zpartition$ correspond to thermodynamics quantities of interest and singularities in these derivatives generally correspond to phase transitions.

The generalization of classical statistical mechanics to quantum statistical is suprisingly straightforward
\begin{equation}
    \zpartition_{\textup{Q}}:= \sum_k e^{-\beta E_k} 
    = \sum_k \mel{E_k}{e^{-\beta \hhat}}{E_k}
= \Tr(e^{-\beta \hhat})
\label{eq:quantum_partition}
\end{equation}
where $\ket{E_k}$ is the eigenstate of the quantum Hamiltonian $\hhat$ associated with its eigenvalue $E_k$.
The last equality in \cref{eq:quantum_partition} is attributed to the fact that trace value is independent of basis.
Hence, we can write the trace as an integral over arbitrary basis $\ket{q}$
\begin{align}
    \zpartition_{\textup{Q}} 
     = \int \dd{q} \mel{q}{e^{-\beta \hhat}}{q}
= \oint \D q(\tau) 
    \exp(-\int_{0}^{\beta }\dd{\tau} 
        \lagrangian_{\textup{E}} (q,\dot{q};\tau)
    )
:= \oint \D q \; e^{-\eaction [q]} 
	\label{eq:partition_path}
\end{align}
the first equality can be identified as the sum over all propagators with \emph{closed} paths in imaginary time interval $\beta$.
The second equality can be derived in the manner of Feynman's path integral cf. \cref{eq:path_integral} with the substitution $t \to -\ii \hbar \tau$ (Wick's rotation)
and $\oint$ emphasizes the periodic boundary condition $q(\beta) = q(0)$.
We should be aware that Lagrangian
$\lagrangian_{\textup{E}} (q,\dot{q};\tau) = \frac{1}{2}m\dot{q}^2(\tau) + V(q(\tau))$ takes the form of classical Hamiltonian which has a sign diffference from real-time action \cref{eq:minkowski_action}.
For this reason, \cref{eq:partition_path} is also called Euclidean (imaginary-time) path integral.

\subsection{Path integral for Ising spins (discrete variable, finite degrees of freedom)}
Although the Hamiltonian of a single spin is quite simple, its path integral formulation is a highly non-trivial problem.
One successful formulation is based on \emph{spin coherent state}. 
Since this formulation is of very rich research interest, e.g., Berry (topological) phase related with adiabatic process,
it deserves a separate paper to discuss its application in quantum computation.
We will not touch it in this paper.
Readers interested in the details are encouraged to read \cite[Sec 3.3.5]{altlandCondensedMatterField2010} based on the paper \cite{stoneSupersymmetryQuantumMechanics1989}.

\subsubsection{Classical-quantum correspondence: Transfer matrix and Suzuki-Trotterization}\label{sec:suzuki}
The path integral for Ising spins that we use in this paper is credited to Suzuki \cite{suzukiQuantumMonteCarlo1993}: inserting the complete set of finite (discrete) configuration basis instead of infinite (continuous) spin coherent states.
Before delving into Suzuki's formulation, it is instructive to see its ``inverse'' process, \emph{transfer matrix method} (Kramers and Wannier (1941)), mapping the scalar description of classical Ising systems to a matrix representation.
Consider a classical $n$ Ising spin (one-dimensional) ring with external field,
the system takes the Hamiltonian $H =-\sum_{l=1}^n \qty(h \sigma^{(l)}\sigma^{(l+1)} + J \sigma^{(l)} )$ with coupling constants $h,J$ and periodic condition $\sigma^{(1)}=\sigma^{(n+1)}$.
Since every Ising spin only takes two values $\qty{\pm 1}$,
every possible nearest-neighbor configuration can be arranged into a 2-by-2 matrix.
Its classical partition function reads
\begin{align}
	\zpartition_{\textup{C}}
	&= \sum_{\sigma^{(1)}\in \qty{\pm 1}}\cdots \sum_{\sigma^{(n)}\in \qty{\pm 1}} 
	\exp(\beta \sum_{l=1}^n \qty(h \sigma^{(l)}\sigma^{(l+1)} +J \sigma^{(l)} ) )
	\\
	&= \sum_{\sigma^{(1)}\in \qty{\pm 1}}\cdots \sum_{\sigma^{(n)}\in \qty{\pm 1}} 
	\mel{\sigma^{(1)}}{\hat{T}}{\sigma^{(n)}} 
	\mel{\sigma^{(n)}}{\hat{T}}{\sigma^{(n-1)}} \cdots
	\mel{\sigma^{(3)}}{\hat{T}}{\sigma^{(2)}} 
	\mel{\sigma^{(2)}}{\hat{T}}{\sigma^{(1)}} 
	\label{eq:classical_ising}
\end{align}
where $\hat{T}$ denotes the transfer matrix with entries
$\mel{\sigma^{(l+1)}}{\hat{T}}{\sigma^{(l)}} = \exp(\beta \qty(h \sigma^{(l)}\sigma^{(l+1)}+ J \sigma^{(l)} ))$.
It is easy to write the matrix explicitly ($\beta$ omitted for simplicity)
\begin{equation}
	\hat{T} 
	= \pmqty{e^{h + J}&e^{-h - J}\\e^{-h+ J}&e^{h - J}}
	= \pmqty{e^{h}&e^{-h}\\e^{-h}&e^{h}} \pmqty{e^{J}&0\\ 0&e^{-J}}
	= \hat{T}_1 \hat{T}_2 
\end{equation}
where $\hat{T}_2$ is clearly equal to $e^{J\pz}$. 
With the help of the matrix exponentiation formula for Pauli matrices, e.g., $e^{-\beta \px}= \cosh(\beta) \pI - \sinh(\beta) \px$,  $\hat{T}_1$ can be written as $C e^{B\px}$ with constants $C$ and $B$.
According to the rules of matrix algebra,
all intermediate spin configurations can be absorbed and the partition function becomes a trace of a matrix, i.e., $\zpartition = \sum_{\sigma^{(1)}} \mel{\sigma^{(1)}}{\hat{T}^n}{\sigma^{(1)}}= \Tr(\hat{T}^n)$.

The above derivation and the forms of $\hat{T}_1$ and $\hat{T}_2$ motivate us to think about the partition function of a quantum spin $\hhat =-h\px -J\pz$.
By taking the procedures analogous to Feynman's path integral \cref{eq:time_slice}, the partition function is approximated by time slicing and product formula 
\begin{equation}
    \zpartition_{\textup{Q}}
= \lim_{M\to \infty} \Tr(e^{\Delta\tau h\px} e^{\Delta\tau J\pz})^M
    = \lim_{M\to\infty} 
\sum_{\qty{\vec{q}}}
    \mel{q^{(0)}}{e^{\deltatau h\px}e^{\deltatau J\pz}}{q^{(M-1)}}
\cdots
    \mel{q^{(1)}}{e^{\deltatau h\px}e^{\deltatau J\pz}}{q^{(0)}}
\label{eq:slice_ising_x_z}
\end{equation}
where $\sum_{\qty{\vec{q}}}=\sum_{q^{(0)}}\sum_{q^{(1)}}\dots\sum_{q^{(M-1)}}$ is sum over all \emph{closed} paths (discrete version of $\oint \D q$) and $q^{(l)}\in \qty{0,1}$.
Note that the complete basis $\qty{\ket{q^{(l)}}}$ inserted is the eigenstates of $\pz^{(l)}$ with eigenvalues $\sigma^{(l)}$. 
So, the tranfer matrix entries are
\begin{equation}
	\mel{q^{(l+1)}}{e^{\Delta\tau h\px}e^{\Delta\tau J\pz}}{q^{(l)}}
	=
	e^{\Delta\tau J\sigma^{(l)}} \mel{q^{(l+1)}}{e^{\Delta\tau h\px}}{q^{(l)}}
= 
	C e^{h' \sigma^{(l+1)} \sigma^{(l)} } e^{\Delta\tau J\sigma^{(l)}} 
\label{eq:ising_x_z}
\end{equation}
where $h'=\frac{1}{2}\log\coth (\deltatau h)$ and $C=\sqrt{\frac{1}{2}\sinh(2\deltatau h)}$.
With Euclidean path integral formulation, we find that the operator $\px$ in the quantum Hamiltonian corresponds a classical Ising chain in the Trotter time direction.
Another important point is that all contributions in the sum of \cref{eq:slice_ising_x_z} are now strictly positive as the number of spin flips is even (periodic condition), but it is not always the case with other basis.
Let's check the Hamiltonian $J\px +h\py$: the entries of the transfer matrix $e^{\Delta\tau h\py}$ read
\begin{align}
\mel{q^{(l+1)}}{e^{\Delta\tau h\py}}{q^{(l)}}
	=
	\cosh(\Delta\tau h)\delta_{q^{(l)},q^{(l+1)}}+\ii \sinh(\Delta\tau h)(q^{(l)}-q^{(l+1)}).
	\label{eq:sign_problem}
\end{align}
With the above equation, it is easy to convince oneself that the phase of the contributions in the sum over all paths now can be $\pm 1$, $\pm \ii$ \cite{vonderlindenQuantumMonteCarlo1992},
which leads to the so-called statistical sign problem in path integral quantum Monte Carlo method.
This single quantum spin Hamiltonian is a minimal example showing that sign problem is basis dependent \cite{hatanoRepresentationBasisQuantum1992}.
Loosely speaking, sign problem is NP-hard \cite{troyerComputationalComplexityFundamental2005} because a good basis, just as spectrum remarked before, is a \emph{priori} unknown.

Let's return to the path integral and take one step further, consider a one-dimensional quantum (homogenous) transverse-field Ising chain described by the Hamiltonain 
$\hhat=-h\sum_{j}^n\px_j-J\sum_{j}^n\pz_{j}\pz_{j+1}$.
With \cref{eq:ising_x_z}, it is obvious to have
\begin{align}
	\mel{\vbq^{(l+1)}}{e^{\deltatau h\sum_j\px_j}e^{\deltatau J\sum_j \pz_{j}\pz_{j+1}}}{\vbq^{(l)}}
= &C^n 
\exp(h' \sum_j \sigma^{(l+1)}_j \sigma^{(l)}_{j}) 
	\exp(\deltatau J \sum_j \sigma^{(l)}_j \sigma^{(l)}_{j+1} )
\label{eq:d_to_d+1_mapping}
\end{align}
where one-dimensional quantum transverse Ising chain is mapped to (1+1)-dimensional classical anisotropic Ising grid (lattice).
A more general argument holds: a $d$-dimensional quantum system corresponds to a $(d+1)$-dimensional classical statistical system.
(This result also applies to the XY model.)
With the identity $(\sigma'-\sigma)^2\equiv 2 - 2\sigma'\sigma$,
the Euclidean action of the quantum Ising chain takes the form
\begin{align}
\eaction\qty[\qty(\vbq_1,\dots,\vbq_n)]
	&\simeq 
	\sum_l \sum_j h'/2 \qty(\sigma^{(l+1)}_{j}-\sigma^{(l)}_{j})^2
	+\deltatau J/2 \qty(\sigma^{(l)}_{j}-\sigma^{(l)}_{j+1})^2 
\label{eq:ising_field}
\end{align}
where the first part can be understood as $\integer_2$ version kinetic energy and the second part can be viewed as the (harmonic) potential energy.

\subsubsection{Path integral for fields (many-body system, infinite degrees of freedom)}\label{sec:path_integral_qft}
The path integral formulation of quantum mechanics is well-suited to the study of systems with an arbitrary number of degrees of freedom. It makes a smooth transition between nonrelativistic quantum mechanics and quantum field theory possible.
However, path integral for fields is a huge topic so that we cannot cover here. 
We only introduce the free field theory as a generalization of \cref{eq:ising_field}.
The common model for illustrating transition from particle to field is a chain of $n$ particles with spacing $a$, connecting the nearest neighbors by springs \cite[Chp1]{altlandCondensedMatterField2010}.
The \emph{Lagrangian density} (Lagrangian of the particle at site $j$) reads
\begin{equation}
	\dlagrangian =
	\frac{m}{2} \dot{q}_j^2 - \frac{k_s}{2} (q_{j+1}-q_j)^2
\end{equation}
where $m$ is the mass of the particle and $k_s$ is the Hooke's spring constant.
The field theory emerges by introducing continuum degrees of freedom $\phi(x)$ and taking continuum limit (lattice spacing $a\to 0$)
\begin{equation}
q_j(t) \to a^{1/2} \eval{\phi(t,x)}_{x=ja}
    ,\quad
    q_{j+1}(t) - q_j(t) \to a^{3/2} \eval{\partial_x \phi(t,x)}_{x=ja}
    ,\quad \sum_{j=1}^n \to \frac{1}{a} \int \dd{x}.
\end{equation}
In the continuum limit, the action,
the space-time integral of the classical Lagrangian density in terms of field and its derivatives
\begin{equation}
	\action_0 [\phi]
= \int\dd{t} \int \dd{x} \dlagrangian_0 (\phi,\partial_x \phi,\dot{\phi})
	,\quad
	\dlagrangian_0 (\phi,\partial_x\phi,\dot{\phi}) =
	\frac{m}{2} \dot{\phi}^2 - \frac{k_s a^2}{2} (\partial_x \phi)^2
\label{eq:lagrangian_qft}
\end{equation}
describes the free scalar field theory.
Note that, in field theory, time and space are on the same footing.
A (0+1)-dimensional quantum field theory is just quantum mechanics.

\section{Complexity Theory}\label{sec:complexity_theory}
This section is written for the readers who are not familiar with the concepts in complexity theory.
In this short appendix, we try to give some intuition of complexity theory rather than formal definitions.
For a fuller account, we refer to \cite{aroraComputationalComplexityModern2009}.
We begin with one of the most important classical complexity classes, NP (nondeterministic polynomial time).
\begin{definition}[NP]\label{def:np}
	Informally, NP comprises languages that can be verified in polynomial time by a deterministic verifier. 
	The celebrated Cook-Levin theorem shows that this class has complete problems, 
	which is in NP and any other language in NP can be reduced to it with polynomial overhead. 
	NP-hard is a class of problems that are informally ``at least as hard as the hardest problems in NP".
	P contains all decision problems that can be solved by a deterministic Turing machine using polynomial time, which refers to the problems efficiently solvable by deterministic classical computation.
\end{definition}
\begin{problem}[$\sat$]\label{def:sat}
	Given a Boolean formula $\Phi (x_1,\cdots,x_{n})$ over $n$ variables,
	$\sat$ is the problem of determining whether there exists a truth assignment (set variables to True or False) such that the Boolean formula is satisfied.
$\sat$ is the first problem that was proven to be NP-complete.
	3$\sat$ is the restricted version of $\sat$ if the formula is an AND of $m$ clauses, i.e.,  $\Phi=c_1\wedge \dots\wedge c_m$ where each clause $c$ is an OR of three variables or their negations.
	($2\sat$ defined similarly) 
	The optimization version of $\sat$, \textsc{MaxSat}, is to find the truth assignment that maximizes the number of clauses satisfied.
\end{problem}
\begin{lemma}
$3\sat$ is NP-complete while $2\sat$ is in P.
\end{lemma}
One may ask how many truth assignments satisfy a given Boolean function. 
We denote this problem as $\#\sat$.
\begin{definition}[$\sharpP$]\label{def:sharp_p}
    Informally, the class of problems counting the number of solutions is $\sharpP$. 
	It is known that $\#\sat$ is $\sharpP$-complete.
\end{definition}
\begin{theorem}
	Evaluating the permanent of a matrix whose entries are 0’s and 1’s is $\sharpP$-complete. 
\end{theorem}
\begin{theorem}
Counting solutions to polynomials over finite fields is $\sharpP$-complete.
\end{theorem}
Calculation of some physical quantities can be reduced to some well-defined computation problems in computer science.
For example, the MaxCut problem is equivalent to minimizing the (classical) Hamiltonian of the Ising model defined on graphs (rather than lattice).
\begin{problem}[MaxCut]\label{prm:maxcut}
    Given a undirected, unweighted graph $G$, 
    $\maxcut$ problem is to partition all vertices into two sets (find the cut)
    such that the number of edges connecting vertices in two sets is maximized.
\end{problem}
\begin{theorem}
	$\maxcut$ (optimization version) is \nameref{def:np}-hard.
	The decision variant (determine whether there is a cut of size at least $k$) is \nameref{def:np}-complete.
\end{theorem}
\begin{theorem}[\cite{barahonaComputationalComplexityIsing1982}]
	Deciding whether the ground state energy of certain classical Ising system is lower than some constant is an NP-complete problem.
    Finding the ground state of a classical Ising-like spin glass in two dimensions is NP-hard.
\end{theorem}
\begin{theorem}[\cite{jerrumPolynomialTimeApproximationAlgorithms1993}]
	Calculating the partition function $\zpartition$ for (classical) frustration systems is a \#P-complete.
	\cite{terhalProblemEquilibrationComputation2000}
\end{theorem}

\begin{problem}[TSP]\label{def:tsp}
    Given a weighted, undirected graph $G$, \emph{traveling salesman problem} (TSP) is to find the Hamiltonian cycle that has the least path (cycle) weight.    
	Hamiltonian cycle of $G$ is a cycle which visits every vertex exactly once.
\end{problem}
\begin{theorem}
	TSP is \nameref{def:np}-hard. The decision version (decide whether the graph has a Hamiltonian cycle of length at most $L$) is NP-complete.
\end{theorem}
\begin{definition}[QMA]\label{def:qma}
Loosely speaking, QMA is the quantum analogue of \nameref{def:np}.
	We don't plan to give the formal definition here.
	For a good introduction to the class QMA, see the book by Kitaev et al. \cite{kitaevClassicalQuantumComputation2002}
	and the paper by Watrous \cite{watrousSuccinctQuantumProofs2000}.
\end{definition}
\begin{theorem}[\cite{kempeComplexityLocalHamiltonian2005}]
    For 2-local quantum Hamiltonians, finding the ground state (quantum analogue of $\sat$) is QMA-hard.
\end{theorem}
\begin{definition}[\BQP]\label{def:bqp}
	The quantum analogue of BPP, $\BQP$ (bounded-error quantum polynomial time) can be informally understood as the class of all languages that is efficiently computable on a Quantum Turing Machine.
\end{definition}

\end{document}